\newif\iffigs\figstrue

\documentclass[12pt]{article}
\usepackage{amstex,epsf,amssymb}
\newtheorem{result}{Result}

\DeclareFontFamily{U}{rsf}{}
\DeclareFontShape{U}{rsf}{m}{n}{
  <5> <6> rsfs5 <7> <8> <9> rsfs7 <10-> rsfs10}{}
\DeclareMathAlphabet\Scr{U}{rsf}{m}{n}

\def\pplogo{\vbox{\kern-\headheight\kern -29pt
\halign{##&##\hfil\cr&{
\ppnumber}\cr\rule{0pt}{2.5ex}&\ppdate\cr}
}}
\makeatletter
\def\ps@firstpage{\ps@empty \def\@oddhead{\hss\pplogo}%
  \let\@evenhead\@oddhead 
}
\def\maketitle{\par
 \begingroup
 \def\thefootnote{\fnsymbol{footnote}}
 \def\@makefnmark{\hbox{$^{\@thefnmark}$\hss}}
 \if@twocolumn
 \twocolumn[\@maketitle]
 \else \newpage
 \global\@topnum\z@ \@maketitle \fi\thispagestyle{firstpage}\@thanks
 \endgroup
 \setcounter{footnote}{0}
 \let\maketitle\relax
 \let\@maketitle\relax
 \gdef\@thanks{}\gdef\@author{}\gdef\@title{}\let\thanks\relax}
\makeatother

\def\Thet{Z}

\def\C{{\mathbb C}}
\def\P{{\mathbb P}}

\def\R{{\mathbb R}}
\def\Z{{\mathbb Z}}

\def\Sl{\operatorname{SL}}
\def\GO{\operatorname{O{}}}

\def\rank{\operatorname{rank}}
\def\Spin{\operatorname{Spin}}
\def\so{\operatorname{\mathfrak{so}}}
\def\su{\operatorname{\mathfrak{su}}}
\def\gu{\operatorname{\mathfrak{u}}}
\def\sp{\operatorname{\mathfrak{sp}}}

\def\CY{Calabi--Yau}

\def\MW{Mordell--Weil}

\def\cM{{\Scr M}}

\def\cD{{\Scr D}}

\def\cL{{\Scr L}}

\def\cX{{\Scr X}}
\def\cG{{\Scr G}}

\def\cMc{{\hfuzz=100cm\hbox to 0pt{$\;\overline{\phantom{X}}$}\cM}}
\def\barcD{{\hfuzz=100cm\hbox to 0pt{$\;\overline{\phantom{X}}$}\cD}}
\def\ff#1#2{{\textstyle\frac{#1}{#2}}}

\def\spnh{\Spin(32)/\Z_2}

\def\HS#1{{\mathbb{F}}_{#1}}
\def\Ist#1{{\mathrm{I}\vphantom{\mathrm{II}}}^*_{#1}}
\def\mf#1{\mathfrak{#1}}
\def\Gloc{\cG_{\mathrm{loc}}}

\begin{document}
\setcounter{page}0
\title{\LARGE Point-like
Instantons on K3 Orbifolds\\[10mm]}
\author{
Paul S. Aspinwall\\[0.7cm]
\normalsize Dept.~of Physics and Astronomy,\\
\normalsize Rutgers University,\\
\normalsize Piscataway, NJ 08855\\[10mm]
David R. Morrison\thanks{On leave from:
Department of Mathematics, Duke University, Durham, NC 27708-0320}\\[0.7cm]
\normalsize Schools of Mathematics and Natural Sciences, \\
\normalsize Institute for Advanced Study, \\
\normalsize Princeton, NJ 08540\\[0mm]
}
\def\ppnumber{\vbox{\baselineskip14pt\hbox{RU-97-29}\hbox{IASSNS-HEP-97/46}
\hbox{hep-th/9705104}}}
\def\ppdate{May, 1997} \date{}

{\hfuzz=10cm\maketitle}

\def\Large{\large}
\def\LARGE{\large\bf}


\begin{abstract}
The map between the moduli space of F-theory (or type II string)
compactifications and heterotic string compactifications can be
considerably simplified by using ``stable degenerations''. We discuss
how this method applies to both the $E_8\times E_8$ and the $\spnh$
heterotic string.  As a simple application of the method we derive
some basic properties of the nonperturbative physics of collections of
$E_8$ or $\spnh$ point-like instantons sitting at $\mathsf{ADE}$
singularities on a K3 surface.
\end{abstract}

\vfil\break


\section{Introduction}

One of the main achievements of the recent progress in the
understanding of string dynamics concerns how nonperturbative physics can
arise in apparently singular situations. For example, if one considers
a type IIA string on a K3 surface, then the conformal field theory
description may break down if the K3 surface acquires orbifold
singularities and an associated component of the $B$-field is set to
zero \cite{me:enhg,w:dyn2}. In this case nonperturbative states, in
the form of gauge particles, become massless rendering the full theory
finite \cite{W:dyn}. That is, nonperturbative gauge symmetries can arise
when the underlying K3 surface becomes singular.

This paper will be concerned with similar effects for the heterotic
string on a K3 surface. In order to compactify the heterotic string,
not only is a compactification space, $S$, required, but also a vector
bundle $E\to S$. The issue of degenerations of the compactification is
thus concerned with degenerations of the bundle structure, as well as
degenerations of $S$. At least in the case where a linear sigma model
may be used to analyze the situation \cite{W:phase}, it is clear that
no breakdown of the conformal field theory is expected if $S$
degenerates while the bundle data remains smooth. Thus, the most
important aspect of analysis of nonperturbative physics in the
heterotic string is the degeneration of the bundle.

The simplest notion of a bundle degeneration is that of the
``point-like instanton'' \cite{W:small-i}. For our purposes an
instanton is an asymptotically flat bundle over $\C^2$ with nonzero
$c_2$ (instanton number) satisfying the Yang-Mills equations of
motion. A point-like instanton is the scaling limit of such a bundle where
all of the curvature has been concentrated into a single point; this
point is then considered to be the location of the instanton. Point-like
instantons (with this local structure) can occur in limits of bundles on
compact surfaces such as the K3 surface.  Several such
instantons may coalesce at the same point in which case the instanton
numbers add.

The simplest notion of a point-like instanton is that in which the
holonomy around the location of the instanton is completely
trivial (and the instanton number is $1$). This leads to two classes
of point-like instanton depending
upon whether we consider the $E_8\times E_8$ or $\spnh$ heterotic
string. Many properties of these instantons on K3 surfaces have
already been analyzed. In the case of the $\spnh$ instanton, the
duality to the type I string may be used to great effect
\cite{W:small-i}, whereas for the $E_8\times E_8$ heterotic string,
M-theory provides an illuminating picture \cite{SW:6d}.

The best tool for analyzing nonperturbative physics is
duality. The duality of interest in this paper relates the heterotic string
on a K3 surface to the F-theory vacuum constructed from some elliptic \CY\
threefold $X$.  One way to view this duality is as a decompactification
limit of a four-dimensional duality along the lines of \cite{KV:N=2}
which relates the type IIA string on $X$ to the heterotic string on a K3
surface times a torus.  One of the most problematic aspects of such dual
pairs is the map between the hypermultiplet moduli space of one theory
to the hypermultiplet moduli space of the other. That is, given a
heterotic string on a certain vector bundle on a certain K3 surface,
what is the corresponding complex structure for $X$?

This problem was greatly simplified by Friedman, Morgan and Witten
\cite{FMW:F} using ``stable degenerations''. We will explain and exploit that
technique in this paper. Not only does it promise to help considerably in
solving the general problem of mapping the moduli space of heterotic
strings to the moduli space of F-theory, it is particularly
well-suited to analysis of point-like instantons. We will thus solve
the simplest problem with this method in this paper.

One of the characteristics of string duality is that there can be many
ways to solve a given problem. The duality between the $\spnh$
heterotic string and the open superstring can allow one to use
alternative methods for analyzing point-like instantons as discussed in
\cite{W:small-i}. The case of point-like $\spnh$ instantons on cyclic quotient
singularities has been analyzed in \cite{In:RG6} using the results of
\cite{DM:qiv} along these lines. While writing this paper we
became aware of \cite{BI:6d1,BI:6d2} which also analyzes this problem
(including the non-abelian quotient singularities) using
orientifolds. In all cases the results agree with the
F-theory analysis below. It is not clear if methods related to open
strings can be applied easily to the point-like $E_8$ instantons.

In section \ref{s:fp} we will quickly review the main tenets of
F-theory we will need. In section \ref{s:stab} we will then analyze
the stable degenerations we require to study the hypermultiplet moduli
spaces. The $E_8$ instantons will then be analyzed in
section \ref{s:E} and the $\spnh$ instantons in section \ref{s:D}. Some
connections between these two cases will be drawn in section
\ref{s:equiv}. We close with some final remarks in section \ref{s:conc}.


\section{The F-theory Picture}  \label{s:fp}

\subsection{F-Theory Models} \label{ss:fm}

In this section we will quickly review the dual F-theory picture which we
will use to study point-like instantons. We refer the reader
to \cite{Vafa:F,MV:F,MV:F2,me:lK3} for further details.

F-theory may be viewed as originating from either the type IIA string (in a
decompactification limit) or the type IIB string (in an unconventional
vacuum with variable dilaton).  Either way,
we begin with a Calabi--Yau manifold $\widehat{X}$ which has an elliptic
fibration $\widehat{\pi}:\widehat{X}\to Z$ with a section, i.e., the image
of a map $\sigma:Z\to\widehat{X}$ such that
$\widehat{\pi}\circ\sigma=\mathrm{id}_Z$. For any such manifold, if we
blow down all components of fibres of $\widehat{\pi}$ which do not
meet the section, then we obtain an elliptic fibration $\pi:X\to Z$ which can
be written in Weierstrass form
\begin{equation}\label{eq:Weq}
  y^2 = x^3 + a(z)x + b(z).
\end{equation}
(The section of the fibration $\pi$ is located at
infinity in the affine coordinates $(x,y)$.)
The blowing down of fibre components of $\widehat{\pi}$ introduces
singularities into $X$, but they are of a relatively mild type known as
canonical singularities.

Let $\cL$ be the inverse of the normal bundle of the section. It is easy to
then show (see, for example, \cite{me:lK3}) that the coefficients $a(z)$ and
$b(z)$ in the Weierstrass equation \eqref{eq:Weq} are sections of
$\cL\,^{\otimes 4}$ and $\cL\,^{\otimes
6}$,  respectively. (Here $z=(z_1,\dots,z_k)$ denotes a coordinate system
on $Z$.)  The
elliptic fibre will be singular over the locus of the zeroes of the
discriminant,
\begin{equation}
  \delta = 4a^3 + 27b^2,
\end{equation}
which is a section of $\cL\,^{\otimes 12}$.
We will use upper case
letters $A$, $B$, and $\Delta$ for the divisor classes in $Z$ corresponding
to $a$, $b$, and $\delta$ and we use $L$ to denote the divisor
associated to the line bundle $\cL$.
The canonical class of $X$ can then be determined to be
\begin{equation}
  K_X = \pi^*(K_Z+L).   \label{eq:KX}
\end{equation}
Thus, to obtain $K_X=0$ we must set $\cL$ equal to the anticanonical line
bundle on $Z$.

The shape of the locus of the
degenerate fibres given by the discriminant, $\Delta$,
is central to the geometry and physics of the F-theory
picture. Each irreducible component of the discriminant within
$\Thet$ may be labeled according to the generic order of vanishing of
$a$, $b$, and $\delta$ along that component. These determine the type of
fibre according to a classification given by Kodaira, displayed in table
\ref{tab:algorithm}.  The generic singularity of $X$ over that component is
a surface singularity of the form $\C^2/G$, as specified in the
table.\footnote{Normal upper case will denote
Lie groups and sans serif will denote finite subgroups of $\Sl(2,\Z)$, or
the corresponding surface singularities.}
The first entry in the table---a Kodaira fibre of type
$\mathrm{I}_0$---corresponds to a divisor on $Z$ which is {\em not}\/ a
component of the discriminant locus, and the last entry in the table
corresponds to a divisor on $Z$ over which $X$ has worse than canonical
singularities (in which case the Weierstrass model is said to be
``non-minimal'').

\begin{table}
\renewcommand{\arraystretch}{1.2}
$$\begin{array}{|ccc|c|c|c|}
\hline
o(a)&o(b)&o(\delta)&\text{Kodaira fibre}&\text{singularity}&\text{gauge
algebra}\\
\hline
\ge0&\ge0&0&\mathrm{I}_0&\text{---}&\text{---}\\
0&0&1&\mathrm{I}_1&\text{---}&\text{---}\\
0&0&2n\ge2&\mathrm{I}_{2n}&\mathsf{A}_{2n-1}&\su(2n)\text{ or }\sp(n)\\
0&0&2n{+}1\ge3&\mathrm{I}_{2n+1}&\mathsf{A}_{2n}&\su(2n{+}1)\text{ or
}\so(2n{+}1)\\
\ge1&1&2&\mathrm{II}&\text{---}&\text{---}\\
1&\ge2&3&\mathrm{III}&\mathsf{A}_1&\su(2)\\
\ge2&2&4&\mathrm{IV}&\mathsf{A}_2&\su(3)\text{ or }\su(2)\\
\ge2&\ge3&6&\Ist0&\mathsf{D}_4&\so(8)\text{ or }\so(7)\text{ or
}\mf{g}_2\\
2&3&n{+}6\ge7&\Ist n&\mathsf{D}_{n+4}&\so(2n{+}8)\text{ or
}\so(2n{+}7)\\
\ge3&4&8&\mathrm{IV}^*&\mathsf{E}_6&\mf{e}_6\text{ or }\mf{f}_4\\
3&\ge5&9&\mathrm{III}^*&\mathsf{E}_7&\mf{e}_7\\
\ge4&5&10&\mathrm{II}^*&\mathsf{E}_8&\mf{e}_8\\
\ge4&\ge6&\ge12&\text{non-minimal}&&\\
\hline
\end{array}$$
\caption{Orders of vanishing, singularities and the gauge algebra}
    \label{tab:algorithm}
\end{table}

In a type IIB construction of the F-theory vacuum corresponding to $X$, the
Weierstrass equation
is used to determine a multi-valued function $\tau(z)$ on the base $Z$ of
the fibration
which specifies the value of the complexified type IIB dilaton; a consistent
vacuum also requires the presence of
Dirichlet $7$-branes wrapped around the components of the discriminant
locus in a manner dictated by the monodromy of the $\tau$ function.  That
monodromy is determined by the Kodaira fibre type, and the corresponding brane
configuration leads to enhanced gauge symmetry.  This story is understood
in full detail only in the cases of $\mathrm{I}_n$ and
$\Ist n$ fibres, which lead to enhanced gauge symmetry based on the classical
groups.

Alternatively, we may regard the F-theory vacuum as a decompactification
limit of the type IIA theory or of M-theory compactified on $X$.  (The
limit taken is the one in which the type IIA string coupling becomes strong and
the area of each elliptic curve becomes small; thanks to the type IIA/M-theory
and M-theory/F-theory dualities, the limiting theory is effectively
decompactified.)  The way
in which the singularities of $X$ lead to enhanced gauge symmetry in those
theories is
understood in detail, and we can expect the same gauge symmetry in the
F-theory limit.  The resulting gauge algebra is shown in the final
column of table \ref{tab:algorithm}.  The different choices of gauge
algebra for each Kodaira fibre are distinguished by the monodromy of the
blown down fibre components as one moves along cycles within the given
component of the discriminant locus.  For details about this distinction
and how to calculate it we refer to \cite{AG:sp32,BKV:enhg,me:lK3}.
Note that there cannot be any monodromy when $X$ is a K3 surface; in the
absence of monodromy we always get one of the simply-laced gauge algebras
$\su(k)$, $\so(2k)$, or $\mf{e}_k$ (the left-most algebra in the final
column of table \ref{tab:algorithm}).

It should be
mentioned that if the \MW\ group of the fibration is of nonzero
rank, that is, if the number of sections of the fibration $\pi$ is
infinite, then there will be additional abelian factors in the gauge
group \cite{MV:F2}. We will largely ignore this possibility but will
mention it briefly in section \ref{s:equiv}. A finite part of the
\MW\ group (if present) can also have interesting effects on the global
form of the gauge group. To avoid having to worry about such
eventualities, we have only concerned ourselves with the gauge {\em
algebras\/} here. The gauge groups are actually
non-simply-connected in many of the examples we discuss below.

Our starting point above was a space $X$ which was obtained from a smooth
Calabi--Yau by a very specific procedure of blowing down certain fibre
components.  This is actually more restrictive than is necessary to make an
F-theory construction.  In fact, any time we have a Weierstrass elliptic
fibration $\pi:X\to Z$ for which $X$ is a Calabi--Yau space with at worst
canonical singularities, the type IIA and M-theory compactifications on $X$
make
sense as limiting theories which occur at finite distance in the moduli
space.  (These theories may involve exotic renormalization group fixed
points in four and five dimensions.)  Taking the F-theory limit of such a
theory is also sensible; the result will often involve an exotic fixed
point in six dimensions
\cite{w:dyn2,SW:6d,W:strong,Sei:6d,In:RG6,DFKS:56,BV:eng}.

What distinguishes these more general models is the fact that blowing up
the singularities responsible for the gauge symmetry enhancement does not
fully resolve the singularities of $X$.  The full resolution may require
further blowups,
which in general must also be accompanied by blowups of the base $Z$ of the
fibration.  The physical interpretation of such blowups of the base
is that we have passed to a kind of Coulomb branch of the exotic fixed
point on which nonzero expectation values have been given to new massless
tensors (or the dimensional reductions of such, if the complex dimension of $X$
is greater than $3$).

To summarize briefly, the two key rules of F-theory for a Calabi--Yau
threefold $X$ are:
\begin{enumerate}
  \item A curve of fibres other than $\mathrm{I}_1$ or $\mathrm{II}$
gives a nonabelian gauge symmetry, as in table \ref{tab:algorithm}.
  \item A blowup within the base, required to form a smooth model for
$X$, corresponds to a massless tensor supermultiplet.
\end{enumerate}
It should be emphasized that both rules assume that any Ramond-Ramond moduli
from the type IIA string have been set to zero. Thus we will only
really be studying a slice of the moduli space when we apply such
rules. Some effects of this restriction were discussed in \cite{Sen:F-GP}.

F-theory also allows the spectrum of massless hypermultiplets to be
deduced as discussed in \cite{BKV:enhg} for example.
One can also deduce the hypermultiplet spectrum from anomaly
cancellation in a related way as discussed in \cite{Sad:anom}, or by a study
of singularities \cite{KV:hyp}.
We will not
concern ourselves with these hypermultiplets in this paper as the
results would be somewhat laborious to list.

\subsection{Duality} \label{ss:dual}

At the heart of
heterotic/F-theory duality
lies a series of fibrations which we now discuss.
One way of describing this duality begins with the basic hypothesis that
the type IIA string on a K3 surface is dual to the heterotic string on a
4-torus. If the K3 surface is in the form of an elliptic fibration with at
least one global section, one can pass to the F-theory limit which
effectively decompactifies two of the dimensions.  On the heterotic string,
the corresponding limit is a straightforward decompactification along two
of the circles of the 4-torus.

Note that the moduli space of the heterotic string on a 2-torus is given by
\begin{equation}\label{eq:M1}
  \Scr{M}_1 = \left({}^{\strut}
         \GO(\Gamma_{2,18})\backslash\GO(2,18)/(\GO(2)\times\GO(18))
         \right) \times \R,
\end{equation}
where the final factor is the eight-dimensional dilaton.
Here, $\Gamma_{2,18}$ denotes the unique even unimodular lattice of
signature $(2,18)$, and
$\GO(\Gamma_{2,18})$ is the discrete group of isometries of that lattice.
{}From the F-theory point of view, the first factor in \eqref{eq:M1} is the
moduli space of
complex structures on an elliptic K3 with a section.

Our primary aim in this paper is to analyze the heterotic string on a K3
surface, $S_H$. To
this end, let us assume that $S_H$ is in the form of an elliptic
fibration $\pi_H:S_H\to C$ with a section, where $C$ is a rational curve,
and that an appropriate bundle has been specified on $S_H$.
We then take each elliptic fibre of the fibration given by $\pi_H$, together
with the restriction of our bundle to that fibre, and
replace it by the corresponding K3 surface which is its F-theory
dual. Thus, we
replace the elliptic K3 surface, $S_H$, by a \CY\ threefold, $X$, in the
form of a K3 fibration, $\pi_A:X\to C$. This allows the heterotic
string on $S_H$ to be analyzed in terms of F-theory on $X$.

We may reintroduce the 2-torus we decompactified by observing that the
type IIA string compactified on $X$ is dual to the heterotic string
compactified on $S_H\times T^2$. The fact that this latter $T^2$ may be
decompactified can be used to show that $X$ itself must be in the form
of an elliptic fibration, $\pi_F:X\to\Thet$, with a section, where $\Thet$
is some algebraic surface.

Finally we see that $\Thet$ itself must be in the form of a fibration
$\pi_B:\Thet\to C$, where $\pi_A = \pi_B\circ\pi_F$. The fibres of
the map $\pi_B$ are rational curves. Since $C$ is also a rational curve,
$\Thet$ must be one of the
Hirzebruch surfaces $\HS n$.  Our focus will be on examples in which $X$
has canonical singularities beyond those responsible for the gauge
symmetry, so typically $\HS n$ will be blown up at a number of points when
studying the Coulomb branch.


\section{Stable Degenerations}   \label{s:stab}

In this section we study the eight-dimensional heterotic/F-theory duality
in more detail.
Given that F-theory on a K3 surface is dual to the heterotic string on
an elliptic curve we would like to know exactly which K3 surface
is mapped to which elliptic curve. As a first step we should switch
off all the Wilson lines around the heterotic elliptic curve. This
restores the full primordial gauge group and hence the K3 surface
acquires either two $\mathsf{E}_8$ singularities (i.e.,
$\C^2/\mathsf{E}_8$) or a
$\mathsf{D}_{16}$ singularity.
The remaining moduli space after fixing this data is
\begin{equation}
  \cM_8 = \GO(\Gamma_{2,2})\backslash\GO(2,2)/(\GO(2)\times\GO(2)),
\end{equation}
(suppressing the dilaton factor).
As far as F-theory is concerned this is the moduli space of complex
structures of a class of singular K3 surfaces and as far as the
heterotic string is concerned this is the Narain moduli space of a
2-torus (with trivial Wilson lines).

For the $E_8\times E_8$ case, we may take the K3 surface to be an elliptic
K3 with a section in Weierstrass form \cite{MV:F}
\begin{equation}
  y^2 = x^3 + a_4s^4\,x + (b_5s^5 + b_6s^6 + b_7s^7),      \label{eq:8E}
\end{equation}
where $s$ is an affine coordinate on the base $\P^1$. This puts
$\mathrm{II}^*$ fibres along $s=0$ and $s=\infty$, each forming an
$\mathsf{E}_8$
singularity on the Weierstrass model and leading to an $E_8$ factor in the
gauge group.

The $\spnh$ is a little more subtle. We need a $\mathsf{D}_{16}$ singularity
which is produced by an $\Ist{12}$ fibre. This only guarantees a group
whose algebra is $\so(32)$. To get $\spnh$ on the nose we require an
elliptic K3 with two sections \cite{AG:sp32}. We can obtain this from
a Weierstrass equation of the form\footnote{This is essentially the same as
written in
\cite{AG:sp32} except that we have put the second section along $x=0$
to simplify the analysis slightly.}
\begin{equation}
  y^2 = x^3 + p(s)\,x^2 + \varepsilon\,x,   \label{eq:8D}
\end{equation}
where $p(s)$ is a cubic in $s$ and $\varepsilon$ is a constant. This
puts the $\Ist{12}$ fibre at $s=\infty$.

Up to a couple of $\Z_2$ factors, $\cM_8$ is isomorphic to two copies
of the complex upper half-plane divided by $\Sl(2,\Z)$. This moduli
space is thus parameterized by two complex numbers
$(\tau,\sigma)$---one giving the complex structure of the heterotic
elliptic curve and
the other giving its K\"ahler form and $B$-field. Taking into account
the reparametrizations of the respective elliptic fibrations, both
(\ref{eq:8E}) and (\ref{eq:8D}) can be shown to depend upon two
complex parameters. We should then be able to write down algebraic
relations between these parameters and $(\tau,\sigma)$ to give the
precise map between the F-theory K3 surface and the heterotic elliptic
curve. This was done for the $E_8\times E_8$ string in
\cite{CCLM:8E}. That result is actually more specific than we
require and not in a suitably geometrical form to be handled easily.

In fact, a direct geometric interpretation of the parameter $J(\sigma)$
could not be expected, since the $\Sl(2,\Z)$ action on $\sigma$ makes
non-geometric identifications between elliptic curves of area $A$ and
$1/A$.  However, if we consider the large area limit $\sigma\to i\infty$,
the remaining parameter $\tau$ has a clear geometric interpretation.

Clearly as $\sigma\to i\infty$, the dual K3 surface will approach the
boundary of its moduli space and degenerate in some way. This will not
be a particularly nice degeneration, as we are moving an
infinite distance in the moduli space. In string theory we are
familiar with what happens for degenerations of K3 surfaces which
acquire an orbifold singularity. Such an orbifold degeneration is a
finite distance event, however, so not at all what we expect here.

To handle these more severe degenerations we will use the language of
``stable degenerations'' as introduced into this subject exactly in
this context by Friedman, Morgan and Witten \cite{FMW:F}. There is an
extensive literature on degenerations of K3 surfaces; we refer the
interested reader to \cite{FM:BGOD} for general information on
this subject.

The idea is to consider a family of K3 surfaces in the form
\begin{equation}
   \pi:\cX\to\Delta,
\end{equation}
where $\cX$ is a complex threefold, $\Delta$ is a complex disc with
parameter $t$, and the generic fibre, $\cX_t$, is a smooth K3
surface. The limit $t\to 0$ will be the degeneration we are interested
in. The first simplification is to assume that we have a ``semistable''
degeneration, which means that $\cX$ is
smooth and the central fibre $\cX_0$ is a reduced divisor within $\cX$
with at most normal crossings. (It is always possible to convert a given
degeneration into this form after replacing $t$ by an appropriate $t^k$ and
blowing up and down.)  The next simplification is to assume
that $K_{\cX}=0$, which can again be achieved by blowing up and down
\cite{Kul:deg,Perss:deg}. Under these circumstances, there is a dictionary
which relates the structure of the central fibre to the monodromy (see
\cite{Mor:CS}, for example).  Even after all of these simplifications,
semistable degenerations with
$K_{\cX}=0$ are not uniquely specified by the behavior of the generic
fibre; however, by blowing down further to a ``stable'' degeneration we can
capture the behavior of the family more accurately.

Let us review what happened in \cite{FMW:F} when this was applied to the
$E_8\times E_8$ case. To begin, take the mildly singular K3 surface which
just has two $\mathsf{E}_8$ singularities given by $\mathrm{II}^*$ fibres.
The discriminant locus of the bad fibres is given by
\begin{equation}
\delta = s^{10}\left(4a_4^3s^2 + 27\left(b_5+b_6s+b_7s^2\right)^2\right).
\end{equation}
Assuming the coefficients are generic, the
model for the elliptic fibration of the K3 surface may then be represented
as
\begin{equation}
\setlength{\unitlength}{0.7mm}%
\begin{picture}(100,15)
\thinlines
\put(0,0){\line(1,0){100}}
\put(10,0){\makebox(0,0){$\times$}}
\put(11,6){\makebox(0,0){$\mathrm{II}^*$}}
\put(30,0){\makebox(0,0){$\times$}}
\put(30,5){\makebox(0,0){$\mathrm{I}_1$}}
\put(40,0){\makebox(0,0){$\times$}}
\put(40,5){\makebox(0,0){$\mathrm{I}_1$}}
\put(60,0){\makebox(0,0){$\times$}}
\put(60,5){\makebox(0,0){$\mathrm{I}_1$}}
\put(70,0){\makebox(0,0){$\times$}}
\put(70,5){\makebox(0,0){$\mathrm{I}_1$}}
\put(90,0){\makebox(0,0){$\times$}}
\put(91,6){\makebox(0,0){$\mathrm{II}^*$}}
\put(110,0){\makebox(0,0){$C$.}}
\end{picture}   \label{eq:Eok}
\end{equation}
The degeneration of this at infinite distance (corresponding to $\sigma\to
i\infty$) would be to push the
$\mathrm{I}_1$ fibres into the $\mathrm{II}^*$ fibres. Let us focus on
two $\mathrm{I}_1$ fibres coalescing with a particular $\mathrm{II}^*$
fibre. (It is impossible for only one $\mathrm{I}_1$ fibre to coalesce with
a $\mathrm{II}^*$ fibre.)  We can model this by looking in a
neighbourhood of $s=0$ for the family
\begin{equation}
  y^2 = x^3 + s^4\,x + s^5t.   \label{eq:degE8}
\end{equation}
In the $(s,t)$ plane, we have a picture like
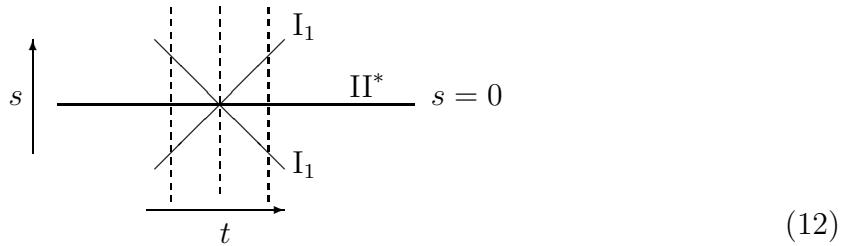
\begin{equation}
\setlength{\unitlength}{0.008500in}%
\begin{picture}(260,134)(50,680)
\thinlines
\put( 80,760){\line( 1, 0){220}}
\put(140,800){\line( 1,-1){ 80}}
\put(220,800){\line(-1,-1){ 80}}
\put( 65,730){\vector( 0, 1){ 70}}
\put(135,695){\vector( 1, 0){ 85}}
\put(180,675){\makebox(0,0)[lb]{$t$}}
\put( 50,760){\makebox(0,0)[lb]{$s$}}
\put(310,760){\makebox(0,0)[lb]{$s=0$}}
\put(260,765){\makebox(0,0)[lb]{$\mathrm{II}^*$}}
\put(225,800){\makebox(0,0)[lb]{$\mathrm{I}_1$}}
\put(225,715){\makebox(0,0)[lb]{$\mathrm{I}_1$}}
\multiput(150,820)(0.00000,-7.74194){16}{\line( 0,-1){  3.871}}
\multiput(210,820)(0.00000,-7.74194){16}{\line( 0,-1){  3.871}}
\multiput(180,820)(0.00000,-7.93103){15}{\line( 0,-1){  3.966}}
\end{picture}   \label{eq:pE8i}
\end{equation}
The three vertical dotted lines do not represent part of the discriminant
locus but rather three different rational curves for the base of
the elliptic K3 as $t$ varies. The vertical line in the middle is the base for
the degenerate K3.
The elliptic threefold, $\cX$, produced by this family has a canonical
singularity worse than the curve of $\mathsf{E}_8$ singularities associated
to the
Weierstrass model of a $\mathrm{II}^*$ fibre.  To resolve this
singularity, a blowup is required in the base
in addition to the blowups within the fibre. We may blow up the point
$(s,t)=(0,0)$ by substituting
\begin{equation}
\begin{split}
  s &\mapsto s_1t_1\\
  t &\mapsto t_1.
\end{split}
\end{equation}
The relevant functions of the Weierstrass form of (\ref{eq:degE8}) then become
\begin{equation}
\begin{split}
  a &= s_1^4t_1^4\\
  b &= s_1^5t_1^6\\
  \delta &= s_1^{10}t_1^{12}(s_1^2+1),
\end{split}  \label{eq:abd1}
\end{equation}
where $t_1=0$ is the locus of the exceptional divisor. Let us use $D$
to denote the class of this exceptional divisor. We see that
$(a,b,\delta)$ vanish to order $(4,6,12)$ respectively along $D$, in other
words, that this is a non-minimal Weierstrass model.  We can make the model
minimal by a change of coordinates (which is one step in resolving the
singularity of the threefold):
\begin{equation}
x=t_1^2x',\qquad y=t_1^3y'.
\end{equation}
This modifies the coefficients in the Weierstrass equation to
\begin{equation}
\begin{split}
a'&=a/t_1^4= s_1^4\\
b'&=b/t_1^6= s_1^5\\
\delta'&=\delta/t_1^{12}= s_1^{10}(s_1^2+1).
\end{split}
\end{equation}
In particular, we must alter the line bundle, replacing $L$ by $L-D$.
Since $D$ adds to the canonical class of $\Thet$ under the blowup, we see
that this
modification of $L$ will also cancel out any change in $K_{\cX}$ produced
by the blowup.

The rational curve given by the dotted line that passed through
the point which was blown up will now intersect the exceptional divisor
at $s_1=\infty$.
This means our picture has now become
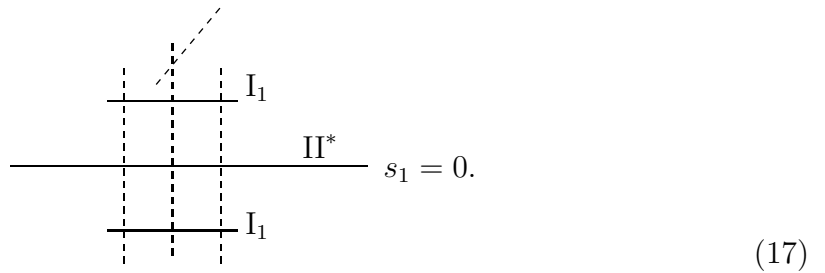
\begin{equation}
\setlength{\unitlength}{0.008500in}%
\begin{picture}(220,155)(80,580)
\thinlines
\put( 80,640){\line( 1, 0){220}}
\put(140,600){\line( 1, 0){ 80}}
\put(220,680){\line(-1, 0){ 80}}
\multiput(150,700)(0.00000,-7.74194){16}{\line( 0,-1){  3.871}}
\multiput(210,700)(0.00000,-7.74194){16}{\line( 0,-1){  3.871}}
\multiput(180,715)(0.00000,-7.87879){17}{\line( 0,-1){  3.939}}
\put(310,630){\makebox(0,0)[lb]{$s_1=0.$}}
\put(260,645){\makebox(0,0)[lb]{$\mathrm{II}^*$}}
\put(225,680){\makebox(0,0)[lb]{$\mathrm{I}_1$}}
\put(225,595){\makebox(0,0)[lb]{$\mathrm{I}_1$}}
\put(170,690){
}
\end{picture}
\end{equation}
The important thing to observe is that at the former location of $t=0$,
the base will now be {\em
two\/} rational curves which intersect at a point. Performing this for
both $\mathrm{II}^*$ fibres, we see that the semistable degeneration
of the picture given by (\ref{eq:Eok}) is the elliptic fibration over
a chain of three rational curves of the form
\begin{equation}
\setlength{\unitlength}{0.7mm}%
\begin{picture}(100,40)
\thinlines
\put(0,35){\line(1,0){100}}
\put(10,40){\line(0,-1){40}}
\put(90,40){\line(0,-1){40}}
\put(10,5){\makebox(0,0){$\times$}}
\put(5,5){\makebox(0,0){$\mathrm{II}^*$}}
\put(10,20){\makebox(0,0){$\times$}}
\put(5,20){\makebox(0,0){$\mathrm{I}_1$}}
\put(10,27){\makebox(0,0){$\times$}}
\put(5,27){\makebox(0,0){$\mathrm{I}_1$}}
\put(90,27){\makebox(0,0){$\times$}}
\put(95,27){\makebox(0,0){$\mathrm{I}_1$}}
\put(90,20){\makebox(0,0){$\times$}}
\put(95,20){\makebox(0,0){$\mathrm{I}_1$}}
\put(90,5){\makebox(0,0){$\times$}}
\put(96,5){\makebox(0,0){$\mathrm{II}^*$}}
\end{picture}   \label{eq:Ess}
\end{equation}
The elliptic fibration over either of the end curves is a rational
elliptic surface whereas the one in the middle is simply a product of
a rational curve and an elliptic curve, i.e., an ``elliptic scroll''.

To recap, our K3 surface has undergone a semistable degeneration into
a chain of three surfaces. The two end surfaces are rational elliptic
surfaces and the middle component is an elliptic scroll. The
intersection between either rational elliptic surface and the elliptic
scroll is the elliptic curve sitting over either intersection point in
(\ref{eq:Ess}). This is an example of a ``type II'' degeneration of a K3
surface \cite{FM:BGOD}.
Note that as there are no bad fibres in the elliptic
scroll part, the elliptic curve must have constant $J$-invariant in
this component.

The elliptic scroll component is somewhat redundant in the description
of the degeneration and we may go to the stable degeneration by
blowing down the middle rational curve in the base (\ref{eq:Ess}).
The base then becomes two rational curves intersecting transversely at
a point.\footnote{One minor complication in this story is that the point in
the base is actually a singular point after this blowdown.}
The K3 itself has become two
rational elliptic surfaces intersecting along an elliptic curve (which
is the fibre over the point of intersection of the two curves in the
base). Let us denote this elliptic curve by $E_*$.
We depict these surfaces in figure \ref{fig:snap}.

\iffigs
\begin{figure}[t]
\begin{center}
\setlength{\unitlength}{0.006250in}%
\begin{picture}(692,290)(60,472)
\thinlines
\put( 60,740){\line( 0,-1){200}}
\put(260,740){\line( 0,-1){200}}
\put(300,640){\vector( 1, 0){ 80}}
\put(501,740){\line(-2,-5){ 62}}
\put(681,740){\line(-1,-3){ 38}}
\put(643,626){\line( 1,-3){ 42}}
\put(440,651){\line( 2,-5){ 60}}
\put(739,620){\vector(-1, 1){ 61}}
\put(740,619){\vector(-4,-3){ 56}}
\put(419,539){\vector( 1, 1){ 61}}
\put( 60,740){
}
\put( 60,540){
}
\put(500,740){
}
\put(500,500){
}
\put(441,650){
}
\put(439,585){
}
\put(453,620){
}
\put(120,472){\makebox(0,0)[lb]{{\scriptsize K3 surface}}}
\put(310,655){\makebox(0,0)[lb]{{\scriptsize $\tau\to i\infty$}}}
\put(752,612){\makebox(0,0)[lb]{{\scriptsize Rational surfaces}}}
\put(390,528){\makebox(0,0)[lb]{$E_*$}}
\end{picture}
\end{center}
  \caption{The stable degeneration.}
  \label{fig:snap}
\end{figure}
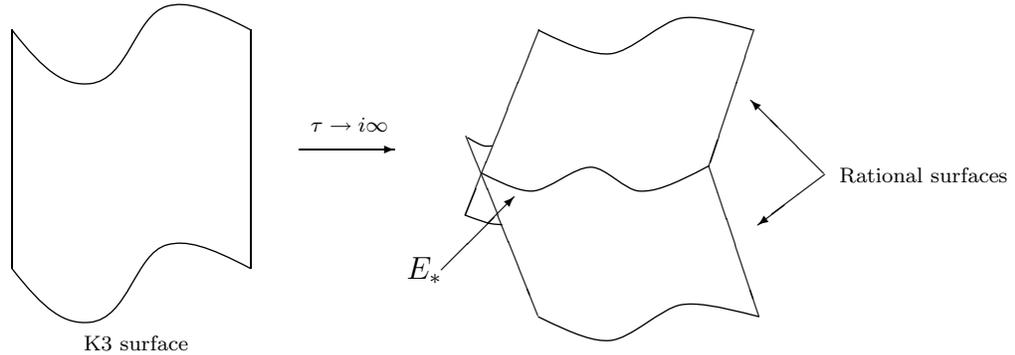
\fi

Now that we have identified the stable degeneration, we can see that there
is actually a more ``fundamental'' way to produce it:  start with a
base surface with a family $\Gamma_\tau$, $\tau\in\C$, of disjoint rational
curves which breaks into
two curves at $\tau=0$, and build a Weierstrass model directly over that
base with $\mathrm{II}^*$ fibres along two disjoint curves transverse to
$\Gamma_\tau$.   If we then set $\tau=t^2$ and resolve singularities, we
get the model with three components which we found above (and could blow
down to the na\"{\i}ve limit in which two $\mathrm{I}_1$ fibres coalesced
with each $\mathrm{II}^*$ fibre).

The complex structure of this stable degeneration of the K3 surface is
not fixed. Although we sent $b_5$ and $b_7$ to zero in (\ref{eq:8E}),
we are still free to vary $a_4$ and $b_6$. These parameters will
control the complex structure of $E_*$. In the heterotic string
interpretation, we
have sent the area of the elliptic curve to infinity but we should be
left with its complex structure as a modulus which we can vary. Clearly the
only way we can consistently map the moduli space of F-theory
compactifications to the heterotic string picture is to identify the
$J$-invariant of $E_*$ with the $J$-invariant of the heterotic
elliptic curve.

Friedman, Morgan and Witten actually go a bit further in this analysis of
the moduli space.  By flopping $\cX$, one may replace the rational elliptic
surfaces above by del Pezzo surfaces. Deforming away the $\mathsf{E}_8$
singular
points within these surfaces then allows the moduli space of the vector
bundle in the heterotic string to be identified \cite{FMW:F}. (See also
\cite{BJPS:F,Don:F} for analysis of the vector bundle moduli space.)

As we shall see, such a clear geometric representation of the
heterotic elliptic curve in the F-theory picture will be of great use
when we consider compactifications on higher dimensional spaces.
We now present an analogue of the first part of the picture
for the $\spnh$
heterotic string, which will be equally useful in our applications.

To obtain the infinite-distance degeneration of the elliptic K3 surface
given by
(\ref{eq:8D}) which corresponds to $\sigma\to i\infty$, we will put
$\varepsilon=t$ and let $t\to0$. That is, we consider the family
\begin{equation}
  y^2 = x^3 + p(s)\,x^2 + t\,x.   \label{eq:8D2}
\end{equation}
This gives rise to the following discriminant in the elliptic
fibration
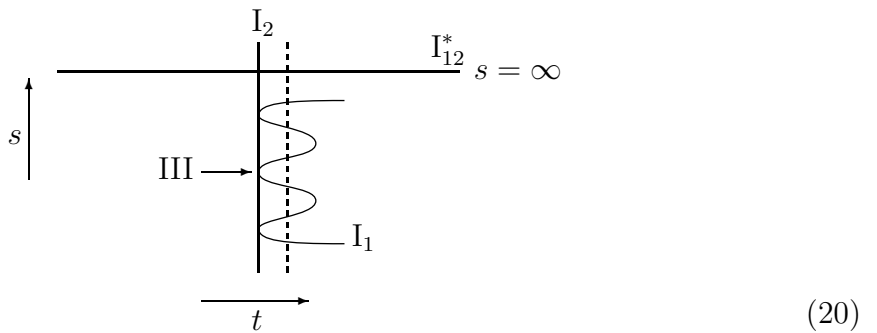
\begin{equation}
\setlength{\unitlength}{0.007500in}%
\begin{picture}(325,214)(45,625)
\thinlines
\put( 80,800){\line( 1, 0){280}}
\put(220,820){\line( 0,-1){160}}
\put( 60,725){\vector( 0, 1){ 70}}
\put(180,640){\vector( 1, 0){ 75}}
\multiput(240,820)(0.00000,-7.80488){21}{\line( 0,-1){  3.902}}
\put(180,730){\vector( 1, 0){ 35}}
\put(280,780){
}
\put(370,795){\makebox(0,0)[lb]{$s=\infty$}}
\put( 45,750){\makebox(0,0)[lb]{$s$}}
\put(215,620){\makebox(0,0)[lb]{$t$}}
\put(340,805){\makebox(0,0)[lb]{$\Ist{12}$}}
\put(285,675){\makebox(0,0)[lb]{$\mathrm{I}_1$}}
\put(215,825){\makebox(0,0)[lb]{$\mathrm{I}_2$}}
\put(150,725){\makebox(0,0)[lb]{$\mathrm{III}$}}
\end{picture}
\end{equation}
This time no blowups are required in the base. However, to obtain a smooth
Calabi--Yau we must blow up along the curve of $\mathrm{I}_2$ fibres, which
automatically blows up
the three $\mathrm{III}$ fibres as well (and of course we must also blow up
the $\Ist{12}$ fibres which we will ignore for
simplicity). Recall that after blowing up, the $\mathrm{I}_2$ and
$\mathrm{III}$ fibres look like:
\begin{equation}
\setlength{\unitlength}{0.007500in}%
\begin{picture}(280,85)(80,695)
\put(280,780){
}
\put(280,720){
}
\put( 80,720){
}
\put( 80,770){
}
\put(110,695){\makebox(0,0)[lb]{$\mathrm{I}_2$}}
\put(305,695){\makebox(0,0)[lb]{$\mathrm{III}$,}}
\end{picture}
\end{equation}
where each curve represents a rational curve.

We see then that at $t=0$ our K3 surface has become a rather peculiar
fibration where {\em none\/} of the fibres are elliptic. What has
happened is that the K3 surface has become a sum of two
``ruled surfaces''---each being a $\P^1$-bundle over $\P^1$. Over a
generic point in the base there are two points of intersection between
the two $\P^1$'s but
there is only one point over the locations of the three $\mathrm{III}$
fibres (where $p(s)=0$) and also over the location of the $\Ist{12}$
fibre (at $s=\infty$). That is the
curve of intersection is a double cover of the base $\P^1$ branched at
four points---i.e., an elliptic curve.

Just as in the $E_8\times E_8$ case, the K3 surface for the F-theory
model corresponding to the $\spnh$ heterotic string breaks into two
rational
surfaces intersecting along an elliptic curve in the stable
degeneration. Again figure \ref{fig:snap} applies and again we can
identify this elliptic curve with that on
which the heterotic string is compactified.

It is worth contrasting the $E_8\times E_8$ and $\spnh$ cases. For the
$E_8\times E_8$ degeneration it was the base $\P^1$ that snapped into
two pieces leaving the elliptic fibre over the point of intersection
as the significant elliptic curve. In the $\spnh$ degeneration it is
the fibre itself which breaks into two pieces leaving the significant
elliptic curve as the double cover of the base. It is remarkable that the
two heterotic strings act so ``oppositely'' in these degenerations.


\section{The $E_8\times E_8$ Heterotic String on a K3 Surface}   \label{s:E}

\subsection{The basic setup}  \label{ss:basic}

The last section was concerned with the map between F-theory on a K3
surface and the heterotic string on an elliptic curve. In this section
we consider the map between F-theory on a \CY\ threefold, $X$, and the
$E_8\times E_8$ heterotic string on a K3 surface, $S_H$.
We will then analyze the $\spnh$ heterotic string in the same setting
in the following section.
The analysis proceeds by assuming that
$S_H$ is in the form of an elliptic fibration with a section. We
then apply the results of the last section ``fibre-wise'' in the spirit
of \cite{FMW:F}.

The resulting $N=1$ theory in six dimensions is much more prone to
quantum corrections than the previous eight-dimensional case and we
need to be careful in how we formulate
precise statements. Let us think of the heterotic/F-theory duality in this
context as originating from the decompactification of a four-dimensional
duality which relates the type IIA
string on $X$ to the heterotic string on $S_H\times
T^2$. Here the
moduli space of hypermultiplets in the type IIA string is prone to
space-time instanton corrections governed by the dilaton. If the dual
pair is of the conventional type (see, for example, \cite{me:lK3} for
details), then the dilaton of the type IIA string is mapped to the
area of a section of the elliptic fibration on $S_H$.
In the limit where we may ignore these corrections, the area of this
section will go to infinity.
To apply the results of the last section we will also let the area of the
fibre in the elliptic fibration on $S_H$ go to infinity. We will therefore
allow the whole
volume of $S_H$ to go to infinity. Only in this large volume limit will
the results of this section apply exactly.

If all the instantons on our K3 surface are point-like with no local
holonomy, then the
holonomy of the degenerated vector bundle over the K3 surface is
completely trivial. Thus, the full primordial $\mf{e}_8\oplus\mf{e}_8$
gauge algebra from the ten-dimensional heterotic string should be
preserved by the compactification. From the F-theory rules, this
amounts to $\Thet$ having two curves of $\mathrm{II}^*$ fibres, producing
two curves of $\mathsf{E}_8$ singularities within $X$.
As explained in \cite{MV:F}, we should
put one of these curves along the exceptional  section $C_0$ of
$\Thet\cong\HS n$ and the other curve along a section $C_\infty$ that does
not intersect the first.

Let $C_0$ denote the divisor class of the exceptional section within
$\Thet$, and $f$ the class of the $\P^1$ fibre of $\Thet\cong\HS
n$. We then have intersection products $C_0\cdot C_0=-n$, $C_0\cdot f=1$,
and $f\cdot f=0$.
The class of the section $C_\infty$  which does not intersect $C_0$ is then
clearly
$C_0+nf$.  Since $K_\Thet = -2C_0-(2+n)f$, we have from (\ref{eq:KX}) that
\begin{equation}
  L = 2C_0+(2+n)f,
\end{equation}
and therefore
\begin{equation}
\begin{split}
  A &= 8C_0+(8+4n)f\\
  B &= 12C_0 + (12+6n)f\\
  \Delta &= 24C_0 + (24+12n)f.
\end{split}
\end{equation}
The functions $(a,b,\delta)$ vanish to order $(4,5,10)$ at a
$\mathrm{II}^*$ fibre. Let us subtract the contributions of
the two curves of $\mathrm{II}^*$ fibres from $A$, $B$, and $\Delta$
to yield
\begin{equation}
\begin{split}
  A' &= 8f\\
  B' &= 2C_0 + (12+n)f\\
  \Delta' &= 4C_0 + (24+2n)f = 2B'.
\end{split}
\end{equation}
Now $B'$ will intersect $C_0$ at $B'\cdot C_0=12-n$ points. Assuming these
points are distinct, and that the intersections are transverse, each
intersection will lead locally to exactly the picture (\ref{eq:pE8i})
which we used in the
stable degeneration. To obtain a smooth model for $X$, we are thus
required to blow up each of these $12-n$ points of
intersection. Similarly, $B'$ collides with the other curve of
$\mathrm{II}^*$ fibres along the divisor $C_\infty=C_0+nf$ a total of
$12+n$ times. We
depict the geometry of the discriminant prior to the
blowup in figure \ref{fig:E8simp}. Note that in this figure the curly
line represents the locus of $\mathrm{I}_1$ fibres. This will be the
case in all subsequent diagrams. The overall shape of this curve is meant to
be only schematic. (In particular, we have omitted the cusps which this
curve invariably has.)  The important aspect is the local geometry of the
collisions
between this curve and the other components of the discriminant which
we try to represent accurately.
\iffigs
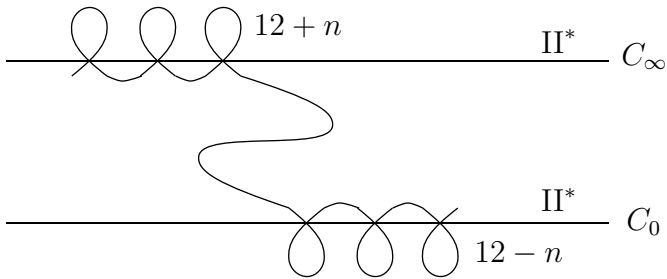
\begin{figure}[t]
\begin{center}
\setlength{\unitlength}{0.008750in}%
\begin{picture}(371,161)(140,651)
\put(309,692){
}
\put(350,692){
}
\put(389,692){
}
\put(330,692){
}
\put(371,692){
}
\thinlines
\put(140,780){\line( 1, 0){360}}
\put(140,683){\line( 1, 0){360}}
\put(179,771){
}
\put(220,771){
}
\put(259,771){
}
\put(200,771){
}
\put(241,771){
}
\put(280,771){
}
\put(460,785){\makebox(0,0)[lb]{$\mathrm{II}^*$}}
\put(460,690){\makebox(0,0)[lb]{$\mathrm{II}^*$}}
\put(288,794){\makebox(0,0)[lb]{$12+n$}}
\put(420,658){\makebox(0,0)[lb]{$12-n$}}
\put(511,676){\makebox(0,0)[lb]{$C_0$}}
\put(508,775){\makebox(0,0)[lb]{$C_\infty$}}
\end{picture}
\end{center}
  \caption{Point-like $E_8$ instantons in the simplest case.}
  \label{fig:E8simp}
\end{figure}
\fi

This is the F-theory picture of the physics discussed in \cite{SW:6d}
that each point-like instanton leads to a massless tensor in six
dimensions (here represented as a blowup of the original base $\HS n$). We
also see that $12-n$ of the instantons are associated
to one of the $E_8$ factors and the other $12+n$ are tied to the other
$E_8$ \cite{MV:F}. After blowing up the base however, one may blow
down in a different way to change $n$. Thus after blowing up, it is not a
well-defined question to ask which $E_8$ a given instanton is associated to.

Now consider what happens to this picture as we go to the stable
degeneration. That is, what happens to the F-theory picture when the
heterotic K3 surface, on which the 24 point-like instantons live,
becomes very large? Along every rational fibre, $f$, of the Hirzebruch
surface, $\HS n$, the process as discussed in section \ref{s:stab}
will occur. That is, every rational fibre will break into two
fibres. Thus our Hirzebruch surface, $\HS n$, will break into two
surfaces which may be viewed as a $(\P^1\vee\P^1)$-bundle over
$\P^1$. The result is shown in figure \ref{fig:E8sdeg}, where $C_*$ is
the curve along which the two irreducible components of the base now
meet. We see that $X$ has broken into two irreducible threefolds
(``generalized Fano threefolds'') which
meet along an elliptic surface with base $C_*$,\footnote{We will
assume that $C_*$ is parallel to the two lines of $\mathrm{II}^*$
fibres as shown in \ref{fig:E8sdeg}. If it is not, we may blow up and
blow down in order to make it so. This is equivalent to reshuffling the
distribution of instantons between the two $E_8$'s.} which is actually a K3
surface as we shall demonstrate below.

\iffigs
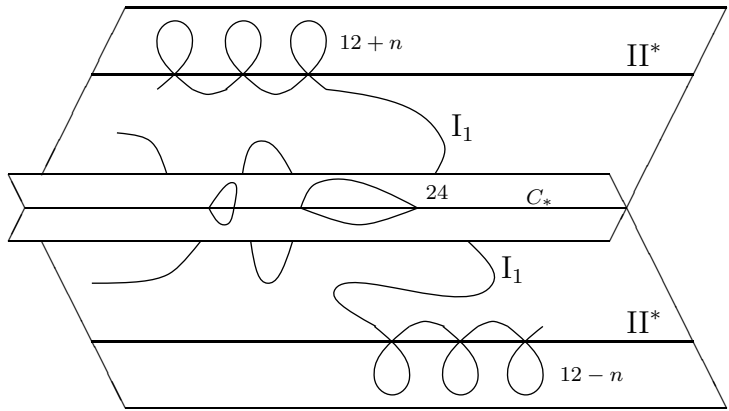
\begin{figure}[htb]
\begin{center}
\setlength{\unitlength}{0.008750in}%
\begin{picture}(430,240)(90,580)
\put(309,629){
}
\put(350,629){
}
\put(389,629){
}
\put(330,629){
}
\put(371,629){
}
\thinlines
\put(140,780){\line( 1, 0){360}}
\put(520,820){\line(-1,-2){ 70}}
\put(450,720){\line( 1,-2){ 70}}
\put(100,700){\line( 1, 0){360}}
\put( 90,720){\line( 1, 0){360}}
\put(160,820){\line(-1,-2){ 50}}
\put( 90,680){\line( 1, 2){ 10}}
\put( 90,720){\line( 1,-2){ 10}}
\put(110,680){\line( 1,-2){ 50}}
\put(160,580){\line( 1, 0){360}}
\put(160,820){\line( 1, 0){360}}
\put(140,620){\line( 1, 0){360}}
\put( 90,680){\line( 1, 0){360}}
\put(179,771){
}
\put(220,771){
}
\put(259,771){
}
\put(200,771){
}
\put(241,771){
}
\put(280,771){
}
\put(335,700){
}
\put(225,700){
}
\put(310,629){
}
\put(185,720){
}
\put(260,720){
}
\put(335,700){
}
\put(260,680){
}
\put(225,700){
}
\put(205,680){
}
\put(288,794){\makebox(0,0)[lb]{{\scriptsize $12+n$}}}
\put(420,595){\makebox(0,0)[lb]{{\scriptsize $12-n$}}}
\put(460,785){\makebox(0,0)[lb]{$\mathrm{II}^*$}}
\put(460,625){\makebox(0,0)[lb]{$\mathrm{II}^*$}}
\put(355,740){\makebox(0,0)[lb]{$\mathrm{I}_1$}}
\put(385,655){\makebox(0,0)[lb]{$\mathrm{I}_1$}}
\put(340,705){\makebox(0,0)[lb]{{\scriptsize 24}}}
\put(400,701){\makebox(0,0)[lb]{{\scriptsize $C_*$}}}
\end{picture}
\end{center}
  \caption{The stable degeneration of point-like $E_8$ instantons.}
  \label{fig:E8sdeg}
\end{figure}
\fi

Before the degeneration, if we had restricted the elliptic fibration
$\pi_F:X\to\Thet$ to one of the rational fibres, $f$, of the
Hirzebruch surface we would have found an elliptic K3 surface. Now
when we look at this elliptic fibration restricted to one of the
$\P^1$'s into which $f$ has broken, we find a rational elliptic surface
instead.
Let us focus on the elliptic fibration, $X_1$, over the lower
component of the surface in figure \ref{fig:E8sdeg}.
Given that the curve, $C_0$, of $\mathrm{II}^*$ fibres was
preserved in this process, this new irreducible component will still have a
section of self-intersection $-n$ and so will still
be $\HS n$. Therefore $C_*=C_0+nf$. Given the
anticanonical class of an elliptic surface is given by the elliptic
fibre, it is not difficult to show that
\begin{equation}
  -K_{X_1} = \pi^*(C_*),
\end{equation}
and so $L=2C_0+(2+n)f-C_*=C_0 +2f$. The class of the discriminant is
then
\begin{equation}
  \Delta = 12C_0 + 24f,
\end{equation}
and so the discriminant passes through $C_*$ a total number of
$\Delta\cdot C_*=24$ times. Generically these will be $\mathrm{I}_1$
fibres. We arrive at the result that the elliptic fibration along
$C_*$ will have 24 $\mathrm{I}_1$ fibres and is therefore a K3
surface. Note that the curve of $\mathrm{I}_1$ fibres from the other
component must pass through $C_*$ at the same 24 points so that the
global geometry makes sense.
We find then that {\em $X_1$ and $X_2$ intersect along a generically
smooth K3 surface}. This K3 surface is, of course, to be identified
with the heterotic string's K3 surface, $S_H$.

We can go one step further in our geometric identification and (partially)
identify the location of the point-like instantons on the K3 surface.  The
F-theory data corresponding to the point-like instantons consists of the
$12\pm n$ intersection points $P_j$ of $B'$ with $C_0$ and $C_\infty$ (at
which $\Delta'$ has double points, as illustrated in many of our figures).
These intersection points are still visible in the stable degeneration,
where each of them lies on some vertical $\P^1$ which meets the curve $C_*$
in a
point $Q_j$.  We assert that the point-like instanton on the heterotic K3
surface $S_H$ will be located at some point along the elliptic curve
$E_j\subset S_H$ which lies over $Q_j\in C_*$.  (The precise locations of
the points within the $E_j$'s will be specified by some of the
Ramond-Ramond moduli.)
This is a very natural geometric prediction; we will provide some specific
evidence for it in section \ref{ss:few}.

\subsection{The $J=0$ series}   \label{ss:J=0}

We may now employ our knowledge of the stable degeneration to answer more
difficult questions about point-like instantons.
This will be a fairly involved process in the general case so we
will start with the
simplest cases. First recall that all of the Kodaira fibres can be
associated with a particular value of the $J$-invariant of the
elliptic fibre, except for $\mathrm{I}_0$ and $\Ist0$ for which $J$ may
take any value (see, for example, page 159 of \cite{BPV:}).

In this section we are going to force a ``vertical'' line of bad fibres
(along an $f$ direction) into the discriminant so that it has a
transverse intersection with the ``horizontal'' line of
$\mathrm{II}^*$ fibres along $C_0$ without any additional local contributions
to the collision from the rest of the discriminant. One may show
\cite{Mir:fibr} that
such intersections of curves within the discriminant must correspond
to fibres with the same $J$-invariant. In this section we require $J=0$
which corresponds to Kodaira types $\mathrm{II}$, $\mathrm{IV}$,
$\Ist0$, $\mathrm{IV}^*$, and $\mathrm{II}^*$. In each case, the
order of vanishing of $\delta$ is twice the order of vanishing of $b$,
with $a$ playing no significant r\^ole. Thus, to analyze the $J=0$ cases we
need only
concern ourselves with the geometry of the divisor $B'$.

For example, let us consider the case illustrated in figure \ref{fig:E8E8}
in which we add a vertical
line of $\mathrm{II}^*$ fibres along the $f$ direction. To do
this, we must subtract $5f$ from $B'$ which implies that what remains can
only produce $7-n$ and $7+n$ simple point-like instantons of the type we
discussed above. It is therefore clear that, whatever else we may have
done to produce this extra line of $\mathrm{II}^*$ fibres, we have had
to ``use up'' ten of the instantons. Note that $B'$ intersects $f$
twice, producing collisions between the $\mathrm{I}_1$ part of the
discriminant and the vertical line of $\mathrm{II}^*$ fibres as shown.
\iffigs
\begin{figure}[htb]
\begin{center}
\setlength{\unitlength}{0.008750in}%
\begin{picture}(371,181)(140,639)
\thinlines
\put(140,780){\line( 1, 0){360}}
\put(140,683){\line( 1, 0){360}}
\put(220,820){\line( 0,-1){180}}
\put(229,699){
}
\put(229,740){
}
\put(229,720){
}
\put(360,771){
}
\put(229,761){
}
\put(259,771){
}
\put(300,771){
}
\put(339,771){
}
\put(280,771){
}
\put(321,771){
}
\put(309,692){
}
\put(350,692){
}
\put(389,692){
}
\put(330,692){
}
\put(371,692){
}
\put(511,676){\makebox(0,0)[lb]{$C_0$}}
\put(508,775){\makebox(0,0)[lb]{$C_\infty$}}
\put(459,784){\makebox(0,0)[lb]{$\mathrm{II}^*$}}
\put(458,689){\makebox(0,0)[lb]{$\mathrm{II}^*$}}
\put(225,639){\makebox(0,0)[lb]{$\mathrm{II}^*$}}
\put(368,794){\makebox(0,0)[lb]{$7+n$}}
\put(420,658){\makebox(0,0)[lb]{$7-n$}}
\end{picture}
\end{center}
  \caption{10 instantons on an $\mathsf{E}_8$ singularity.}
  \label{fig:E8E8}
\end{figure}
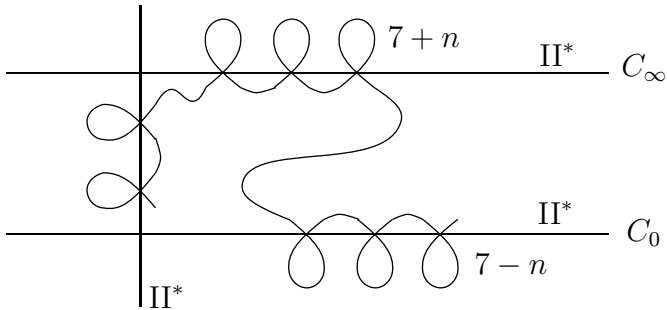
\fi

Now when we consider the stable degeneration of this model, we cannot avoid
having the new line
of $\mathrm{II}^*$ fibres pass through $C_*$. Therefore $S_H$ has an
orbifold singularity of type $\mathsf{E}_8$ (i.e., locally of the form
$\C^2$ divided by
the binary icosahedral group). We claim that this geometry represents 10
point-like instantons sitting on an $\mathsf{E}_8$ quotient singularity in
$S_H$.  This is consistent with our earlier assertion that the vertical
lines determine the location of the point-like instantons; we will discuss
this point more fully in section \ref{ss:few}.

The elliptic fibration of figure \ref{fig:E8E8} is quite singular
and requires many blowups in the base before it becomes smooth. For
example, the degrees of $(a,b,\delta)$ for $\mathrm{II}^*$
fibres are $(4,5,10)$ respectively. Thus, if two such curves intersect
transversely and we blow up the point of intersection, the exceptional
divisor will contain degrees $(8,10,20)$. As in section
\ref{s:stab}, this indicates a non-minimal Weierstrass model, and when
passing to a minimal model, $L$ is adjusted in a way that subtracts
$(4,6,12)$ from these degrees and  restores $K_X$ to $0$.  We are thus
left with an exceptional curve of degrees $(4,4,8)$, which is a curve
of $\mathrm{IV}^*$ fibres. This new curve will intersect the old
curves of  $\mathrm{II}^*$ fibres and these points of intersection
also need blowing up. Iterating this process we finally arrive at smooth
model (i.e., no further blowups need to be done) when we have the chain
\begin{equation}
\setlength{\unitlength}{0.006750in}%
\begin{picture}(800,80)(60,700)
\thinlines
\put( 60,780){\line( 1,-1){ 80}}
\put(180,780){\line( 1,-1){ 80}}
\put(240,700){\line( 1, 1){ 80}}
\put(300,780){\line( 1,-1){ 80}}
\put(360,700){\line( 1, 1){ 80}}
\put(420,780){\line( 1,-1){ 80}}
\put(480,700){\line( 1, 1){ 80}}
\put(540,780){\line( 1,-1){ 80}}
\put(600,700){\line( 1, 1){ 80}}
\put(660,780){\line( 1,-1){ 80}}
\put(780,780){\line( 1,-1){ 80}}
\put(120,700){
}
\put(720,700){
}
\put(105,740){\makebox(0,0)[lb]{{\scriptsize $\mathrm{II}^*$}}}
\put(155,720){\makebox(0,0)[lb]{{\scriptsize $\mathrm{I}_0$}}}
\put(225,740){\makebox(0,0)[lb]{{\scriptsize $\mathrm{II}$}}}
\put(275,720){\makebox(0,0)[lb]{{\scriptsize $\mathrm{IV}$}}}
\put(345,740){\makebox(0,0)[lb]{{\scriptsize $\Ist0$}}}
\put(395,720){\makebox(0,0)[lb]{{\scriptsize $\mathrm{II}$}}}
\put(465,740){\makebox(0,0)[lb]{{\scriptsize $\mathrm{IV}^*$}}}
\put(515,720){\makebox(0,0)[lb]{{\scriptsize $\mathrm{II}$}}}
\put(585,740){\makebox(0,0)[lb]{{\scriptsize $\Ist0$}}}
\put(635,720){\makebox(0,0)[lb]{{\scriptsize $\mathrm{IV}$}}}
\put(705,740){\makebox(0,0)[lb]{{\scriptsize $\mathrm{II}$}}}
\put(755,720){\makebox(0,0)[lb]{{\scriptsize $\mathrm{I}_0$}}}
\put(825,740){\makebox(0,0)[lb]{{\scriptsize $\mathrm{II}^*$.}}}
\end{picture}    \label{eq:E8chain}
\end{equation}
Chains of this sort were studied systematically in
\cite{Mir:fibr,Grs:log}.\footnote{Some care is required in applying the
results of \cite{Mir:fibr} since the
Calabi--Yau condition was not relevant there; for example, a transverse
intersection of a curve of  $\mathrm{II}$ fibres and a curve of
$\mathrm{IV}$ fibres should not be blown up since there is no singularity
in the total space of the fibration.  (This blowup was done for convenience
in \cite{Mir:fibr}, but it is implicit in \cite{Grs:log} that it need not
be done.)}
(Such chains produced by collisions in the discriminant have also been
discussed in the physics literature \cite{BJ:collide,AN:Fc}.\footnote{Note
that the analysis in
\cite{BJ:collide} is erroneous in its assertion that above collision
cannot be resolved by blowups to preserve $K_X=0$.})
We see that in the present example, eleven blowups are required.
Various of the intersections in the above graph produce monodromies
and the usual rules of F-theory \cite{AG:sp32,BKV:enhg} then dictate
that the resulting gauge algebra from this graph will be
\begin{equation}
  \mf{e}_8\oplus\su(2)\oplus\mf{g}_2\oplus\mf{f}_4
  \oplus\mf{g}_2\oplus\su(2)\oplus\mf{e}_8.
\end{equation}
We need two of these chains from the two intersections of curves of
$\mathrm{II}^*$ fibres in figure \ref{fig:E8E8}. Adding this to the 16
further blowups from the $B'$ collisions we obtain our first result.
\begin{result}
  10 point-like $E_8$ instantons on an $\mathsf{E}_8$ quotient singularity
produce 38 extra massless
tensors (in addition to the dilaton) and a gauge algebra
\begin{equation}
  \mf{e}_8^{\oplus3}\oplus\mf{f}_4^{\oplus2}\oplus\mf{g}_2^{\oplus4}\oplus
       \su(2)^{\oplus4}.
\end{equation}
\end{result}

Two of the above $\mf{e}_8$ terms come from the perturbative, primordial
part of the heterotic string. All of the rest is nonperturbative. The
couplings of these nonperturbative components are controlled by
particular massless tensors. Let us introduce $\Gloc$ as the
nonperturbative gauge algebra produced locally by the collision of the
point-like instantons with the quotient singularity. Thus, if we have
a situation where $k$ point-like instantons have coalesced on a
quotient singularity and the remaining $24-k$ instantons are point-like but
lie disjointly on smooth points, the total gauge algebra will be given
by
\begin{equation}
  \cG \cong \mf{e}_8 \oplus \mf{e}_8 \oplus \Gloc.
\end{equation}
Also, let the number of massless tensors be $n_T+1$ (the extra one is
for the dilaton). Then define $n_T'$ by $n_T=n_T'+24-k$. Thus $n_T'$
counts the massless tensors given locally by the instantons on the
quotient singularity. Our ten point-like $E_8$ instantons on an $\mathsf{E}_8$
quotient singularity then give
\begin{equation}
\begin{split}
\Gloc&\cong\mf{e}_8\oplus\mf{f}_4^{\oplus2}
   \oplus\mf{g}_2^{\oplus4}\oplus\su(2)^{\oplus4}\\
n_T' &= 24.
\end{split}
\end{equation}

Now that we have seen how to handle a particular example, we should
try to generalize our techniques. The next question we shall ask is
what happens when there are more than 10 instantons on an $\mathsf{E}_8$
singularity. Clearly, from figure \ref{fig:E8E8}, we may bring one of
the 14 remaining instantons into the $\mathsf{E}_8$ singularity by bringing one
of the remaining intersections of $B'$ with $C_0$ into the collision
of $C_0$ with our vertical line of $\mathrm{II}^*$ fibres.
Fixing this point as $(s,t)=(0,0)$, locally
such a collision will now look something like
\begin{equation}
\begin{split}
  a &= s^4t^4\\
  b &= s^5t^5(s+\alpha t)\\
  \delta &= s^{10}t^{10}\left(4s^2t^2 + 27\left(s+\alpha t\right)^2\right),
\end{split}
\end{equation}
for some constant $\alpha$. We show the form of the discriminant in
the upper half of figure \ref{fig:E8E8E8}. Now when we blow up the
point $(s,t)=(0,0)$, we find that the exceptional curve produced
carries $\mathrm{II}^*$ fibres. We show this in the lower half of
figure \ref{fig:E8E8E8}. We then see that having 11 point-like
instantons on an $\mathsf{E}_8$ quotient singularity is very similar to
the case of 10 point like instantons except that our chain of curves
of $\mathrm{II}^*$ fibres is now longer by one. Thus, the chain
(\ref{eq:E8chain}) will now appear three times rather than just twice.

\iffigs
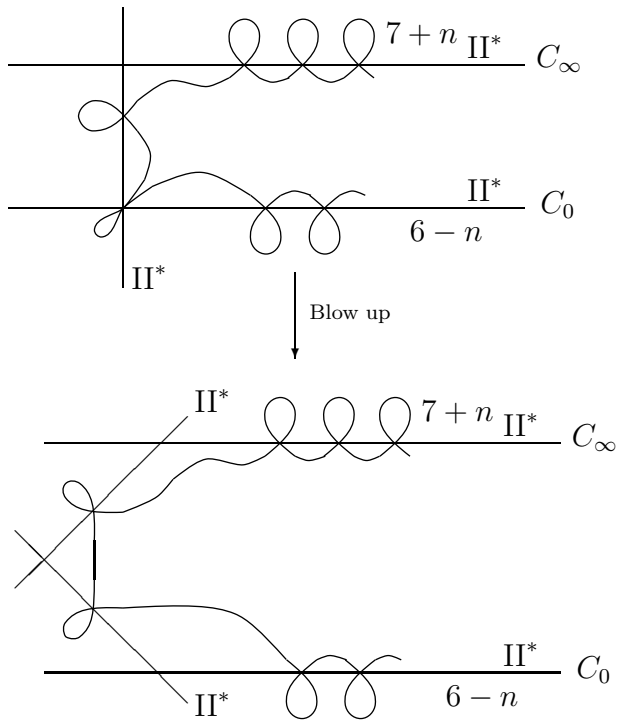
\begin{figure}[tbh]
\begin{center}
\setlength{\unitlength}{0.007500in}%
\begin{picture}(416,496)(140,324)
\thinlines
\put(165,356){\line( 1, 0){360}}
\put(334,365){
}
\put(375,365){
}
\put(355,365){
}
\put(396,365){
}
\put(485,361){\makebox(0,0)[lb]{$\mathrm{II}^*$}}
\put(536,349){\makebox(0,0)[lb]{$C_0$}}
\put(445,331){\makebox(0,0)[lb]{$6-n$}}
\put(165,516){\line( 1, 0){360}}
\put(265,535){\line(-1,-1){120}}
\put(265,335){\line(-1, 1){120}}
\put(319,507){
}
\put(360,507){
}
\put(399,507){
}
\put(340,507){
}
\put(381,507){
}
\put(255,491){
}
\put(533,511){\makebox(0,0)[lb]{$C_\infty$}}
\put(485,521){\makebox(0,0)[lb]{$\mathrm{II}^*$}}
\put(428,530){\makebox(0,0)[lb]{$7+n$}}
\put(340,635){\vector( 0,-1){ 60}}
\put(140,780){\line( 1, 0){360}}
\put(220,820){\line( 0,-1){195}}
\put(140,680){\line( 1, 0){360}}
\put(200,448){\line( 0,-1){ 27}}
\put(309,689){
}
\put(350,689){
}
\put(330,689){
}
\put(371,689){
}
\put(294,771){
}
\put(335,771){
}
\put(374,771){
}
\put(315,771){
}
\put(356,771){
}
\put(235,700){
}
\put(240,695){
}
\put(230,755){
}
\put(235,700){
}
\put(230,755){
}
\put(200,448){
}
\put(220,401){
}
\put(220,468){
}
\put(220,401){
}
\put(350,600){\makebox(0,0)[lb]{\scriptsize Blow up}}
\put(508,775){\makebox(0,0)[lb]{$C_\infty$}}
\put(460,785){\makebox(0,0)[lb]{$\mathrm{II}^*$}}
\put(460,685){\makebox(0,0)[lb]{$\mathrm{II}^*$}}
\put(511,673){\makebox(0,0)[lb]{$C_0$}}
\put(226,624){\makebox(0,0)[lb]{$\mathrm{II}^*$}}
\put(403,794){\makebox(0,0)[lb]{$7+n$}}
\put(420,655){\makebox(0,0)[lb]{$6-n$}}
\put(270,538){\makebox(0,0)[lb]{$\mathrm{II}^*$}}
\put(270,326){\makebox(0,0)[lb]{$\mathrm{II}^*$}}
\end{picture}
\end{center}
  \caption{11 instantons on an $\mathsf{E}_8$ singularity.}
  \label{fig:E8E8E8}
\end{figure}
\fi

By putting $b=s^5t^5(s+\alpha t^\ell)$, for $\ell>1$, we can bring more
instantons into the $\mathsf{E}_8$ singularity and each instanton will
have the effect of increasing the length of the chain of
$\mathrm{II}^*$ fibres by one. This leads to the following
\begin{result}
  A collection of $k$ point-like $E_8$ instantons (for $k\geq 10$)
on an $\mathsf{E}_8$ quotient singularity yields
\begin{equation}
\begin{split}
  \Gloc&\cong\mf{e}_8^{\oplus(k-9)}\oplus\mf{f}_4^{\oplus(k-8)}\oplus
     \mf{g}_2^{\oplus(2k-16)}\oplus\su(2)^{\oplus(2k-16)}\\
n_T'&=12k-96.
\end{split}
\end{equation}
\end{result}

We may follow this same method for analyzing a vertical line of
$\mathrm{IV}$, $\Ist0$, or $\mathrm{IV}^*$ fibres replacing the
vertical line of $\mathrm{II}^*$ fibres in figure \ref{fig:E8E8}.
These produce $\mathsf{A}_2$ (i.e., $\C^2/\Z_3$), $\mathsf{D}_4$, or
$\mathsf{E}_6$ singularities
on the K3 surface, $S_H$, respectively. The other $J=0$ case, namely a
fibre of type $\mathrm{II}$, produces no singularity on $S_H$ and no
interesting nonperturbative physics.

\begin{table}
\renewcommand{\arraystretch}{1.5}
$$
\begin{array}{|c|c|c|l|}
\hline
G&&n_T'&\multicolumn{1}{c|}{\Gloc}\\
\hline
\mathsf{A}_2&k=4&k&\su(2)\\
\mathsf{A}_2&k\geq5&k&\su(2)\oplus\su(3)^{\oplus(k-5)}\oplus\su(2)\\
\mathsf{D}_4&k=6&k&\su(2)\oplus\mf{g}_2\oplus\su(2)\\
\mathsf{D}_4&k\geq7&2k-6&\su(2)\oplus \mf{g}_2\oplus\so(8)^{\oplus(k-7)}
                     \oplus \mf{g}_2\oplus\su(2)\\
\mathsf{E}_6&k=8&10&\su(2)\oplus\mf{g}_2\oplus \mf{f}_4
  \oplus \mf{g}_2\oplus\su(2)\\
\mathsf{E}_6&k\geq9&4k-22&
{\renewcommand{\arraystretch}{1.0}\arraycolsep=0pt
\begin{array}[t]{l}
  \su(2)\oplus \mf{g}_2\oplus \mf{f}_4
  \oplus\su(3)\oplus\bigl(\mf{e}_6\oplus \su(3)\bigr)^{\oplus(k-9)}\\
  \qquad\oplus \mf{f}_4\oplus \mf{g}_2\oplus\su(2)\end{array}}\\
\mathsf{E}_8&k\geq10&12k-96&\mf{e}_8^{\oplus(k-9)}\oplus
\mf{f}_4^{\oplus(k-8)}\oplus
\mf{g}_2^{\oplus(2k-16)}\oplus\su(2)^{\oplus(2k-16)}\\
\hline
\end{array}$$
\caption{$k$ point-like $E_8$ instantons on a $\C^2/G$ singularity for $J=0$.}
    \label{tab:E81}
\end{table}

In table \ref{tab:E81}, we show the result of allowing $k$ point-like
$E_8$ instantons to coalesce on a singularity of type $\C^2/G$.
The local contribution to the number of massless tensors and to the
gauge algebra is listed assuming a given bound on $k$.

\subsection{Fewer instantons}   \label{ss:few}

In each entry in table \ref{tab:E81} imposing the vertical line of bad
fibres forced a minimum number of instantons into the quotient
singularity. How do we analyze the situation when there are fewer
instantons within the singularity?

Let us begin with the $\mathsf{E}_8$ quotient singularity in $S_H$ again.
Consider the elliptic fibration given by the lower half of figure
\ref{fig:E8sdeg} {\em after\/} the stable degeneration has occurred. In
particular we are interested in $B'$, the divisor associated to
$b$ after the contribution from the line of $\mathrm{II}^*$ fibres
along $C_0$ has been subtracted. The important point is that the
class of $B'$ is given by
\begin{equation}
  B' = C_0 + 12f.
\end{equation}
That is, $B'\cdot f=1$, and so $B'$ is the class of a section of $\HS n$.

We have asserted above that a single point-like instanton is associated to
a transverse
intersection of $B'$ with $C_0$. An $\mathsf{E}_8$ singularity in $S_H$
requires a $\mathrm{II}^*$ fibre at a point in $C_*$, which means
that $B'\cdot C_*=5$ at that point. So long as this point within $C_*$ does
not lie exactly vertically above any of the point-like instantons in
$C_0$, we may arrange for $B'$ to be an irreducible curve. This is
shown in figure \ref{fig:E8none}. Note that the solid line in the
figure represents $B'$ and {\em not\/} the discriminant. We are free to
move the locations of all the instantons by varying $B'$ and no
nonperturbative physics is associated with the $\mathsf{E}_8$ singularity in
$S_H$. Clearly this represents zero instantons on the $\mathsf{E}_8$
singularity.

\iffigs
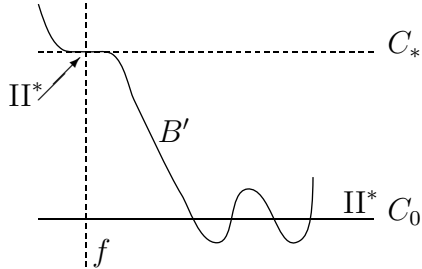
\begin{figure}[htb]
\begin{center}
\setlength{\unitlength}{0.006250in}%
\begin{picture}(295,220)(70,600)
\thinlines
\multiput( 80,780)(7.88732,0.00000){36}{\line( 1, 0){  3.944}}
\put( 80,640){\line( 1, 0){280}}
\multiput(120,820)(0.00000,-8.00000){28}{\line( 0,-1){  4.000}}
\put( 80,740){\vector( 1, 1){ 35}}
\put( 80,820){
}
\put(372,775){\makebox(0,0)[lb]{$C_*$}}
\put(372,635){\makebox(0,0)[lb]{$C_0$}}
\put(335,645){\makebox(0,0)[lb]{$\mathrm{II}^*$}}
\put(125,600){\makebox(0,0)[lb]{$f$}}
\put(180,705){\makebox(0,0)[lb]{$B'$}}
\put( 55,735){\makebox(0,0)[lb]{$\mathrm{II}^*$}}
\end{picture}
\end{center}
  \caption{No instantons on an $\mathsf{E}_8$ singularity.}
  \label{fig:E8none}
\end{figure}
\fi

Now let us bring one of the points of intersection of $B'$ with $C_0$
into the same vertical $f$-line as the quintuple point of intersection
of $B'$ with $C_*$. The only way we may preserve $B'\cdot f=1$ is to let
$B'$ become a sum $B_1+B''$ where $B_1$ lies entirely along the
$f$-line and $B''$ remains a section of $\HS n$. That is we pick up a
line of bad fibres---of type $\mathrm{II}$ in this case. We see then that we
are forced into the geometry discussed in the last section of a
vertical line of bad fibres when the instantons coalesce onto the
orbifold point in $S_H$.
This bolsters our assertion that the vertical fibres serve to tie the
location of the point-like instantons on $S_H$ to the intersection points
of $B'$ with $C_0$.

We should remember that we are dealing with only half the picture
here. We should also worry about the upper half of the stable
degeneration in figure \ref{fig:E8sdeg}. If we have a vertical line of
bad fibres {\em prior\/} to the degeneration then both components will
have the corresponding vertical line after the degeneration and point-like
instantons are required in the orbifold point from both $E_8$'s. Thus,
we are forced to get a vertical line of $\mathrm{II}$ fibres when {\em
two\/} instantons sit at the orbifold point. Similarly we get a
vertical line of $\mathrm{IV}$ fibres when 4 instantons coalesce on
the orbifold point, and so on.

Thus we see that fewer than 10 instantons on a $\mathsf{E}_8$ quotient
singularity reduces to the other cases listed in table
\ref{tab:E81}. Note that when the number of instantons is odd the
two surfaces in the degeneration will end up with different vertical
lines of bad fibres. To determine the nonperturbative physics, we only
care about
the form of the vertical line {\em prior\/} to the degeneration. We
list the results for the $\mathsf{E}_8$ singularity in table \ref{tab:E8low}.

\begin{table}
\renewcommand{\arraystretch}{1.5}
$$
\begin{array}{|c|c|l|}
\hline
k&n_T'&\multicolumn{1}{c|}{\Gloc}\\
\hline
<4&k&-\\
4&4&\su(2)\\
5&5&\su(2)\oplus\su(2)\\
6&6&\su(2)\oplus \mf{g}_2\oplus\su(2)\\
7&8&\su(2)\oplus \mf{g}_2\oplus\mf{g}_2\oplus\su(2)\\
8&10&\su(2)\oplus\mf{g}_2\oplus \mf{f}_4
  \oplus \mf{g}_2\oplus\su(2)\\
9&14&
  \su(2)\oplus \mf{g}_2\oplus \mf{f}_4\oplus\su(3)
  \oplus \mf{f}_4\oplus \mf{g}_2\oplus\su(2)\\
\hline
\end{array}$$
\caption{A few point-like $E_8$ instantons on a $\C^2/\mathsf{E}_8$
singularity.}
    \label{tab:E8low}
\end{table}

Similarly the other $\mathsf{A}_2$, $\mathsf{D}_4$, and $\mathsf{E}_6$
singularities with a small
number of instantons reduce to the results for less singular points
listed in table \ref{tab:E81}.

\subsection{The other singularities}   \label{ss:other}

So far in this section we have only concerned ourselves with the
Kodaira fibres corresponding to $J=0$. This leaves many possible
singularities for $S_H$ unanalyzed. The methods from above may be
employed equally well in this situation but the analysis becomes a
little more complex due the geometry of the discriminant being
dictated by the zeroes of $a$ in addition to those of $b$.

We begin with the case of a vertical line of $\mathrm{I}_m$ fibres
corresponding to a $\C^2/\Z_m$ orbifold singularity in $S_H$. We may
set this situation up by defining
\begin{equation}
\begin{split}
a &= 3s^4(-1+t^m)\\
b &= 2s^5(s+t^\ell)\\
\delta &= 108s^{10}t^m(3s^2 - 3s^2t^m + s^2t^{2m} + 2st^{\ell-m} +
t^{2\ell-m}),
\end{split}  \label{eq:E8Am}
\end{equation}
where we assume that $\ell\geq m$. As required this gives a line of
$\mathrm{II}^*$ fibres along $C_0$, i.e., $s=0$, and a line of
$\mathrm{I}_m$ fibres along $t=0$.

The intersection number of $B'=\{s+t^\ell=0\}$ with $C_0$ is $\ell$ and so
this represents $\ell$ point-like instantons, not including the
contribution from the other $E_8$ horizontal line along $C_\infty$.

The point $s=t=0$ must be blown up to resolve the \CY\ threefold,
$X$. If we assume that $\ell>m$ then the result of this blowup is to
introduce a new exceptional curve of $\mathrm{I}_m$ fibres in addition
to the original. The proper transform of the discriminant in
(\ref{eq:E8Am}) will have $\ell$ lowered by one. We can continue this
process until $\ell=m$. The next time we blow up, the exceptional curve will
be a line of $\mathrm{I}_{m-1}$ fibres, and so on. We are finished when we
finally produce a line of $\mathrm{I}_0$ fibres. We depict this in figure
\ref{fig:E8Am}.

\iffigs
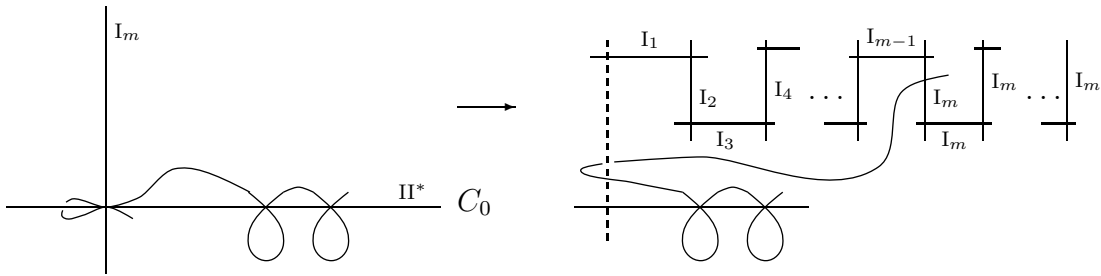
\begin{figure}[tbh]
\begin{center}
\setlength{\unitlength}{0.008750in}%
\begin{picture}(640,160)(20,660)
\put(165,709){
}
\put(186,709){
}
\put(204,709){
}
\put(425,709){
}
\put(446,709){
}
\put(464,709){
}
\thinlines
\put(530,740){\line( 0, 1){ 60}}
\put(525,790){\line( 1, 0){ 50}}
\put(570,800){\line( 0,-1){ 60}}
\put(565,750){\line( 1, 0){ 45}}
\put(605,740){\line( 0, 1){ 60}}
\put(655,800){\line( 0,-1){ 60}}
\put(500,765){\makebox(0,0)[lb]{$\ldots$}}
\put(610,770){\makebox(0,0)[lb]{\scriptsize $\mathrm{I}_m$}}
\put(580,735){\makebox(0,0)[lb]{\scriptsize $\mathrm{I}_m$}}
\put(575,760){\makebox(0,0)[lb]{\scriptsize $\mathrm{I}_m$}}
\put(630,765){\makebox(0,0)[lb]{$\ldots$}}
\put(660,770){\makebox(0,0)[lb]{\scriptsize $\mathrm{I}_m$}}
\put(535,795){\makebox(0,0)[lb]{\scriptsize $\mathrm{I}_{m-1}$}}
\put(290,760){\vector( 1, 0){ 35}}
\put(360,700){\line( 1, 0){140}}
\put( 20,700){\line( 1, 0){260}}
\put( 80,820){\line( 0,-1){160}}
\multiput(380,800)(0.00000,-7.74194){16}{\line( 0,-1){  3.871}}
\put(370,790){\line( 1, 0){ 70}}
\put(430,800){\line( 0,-1){ 60}}
\put(420,750){\line( 1, 0){ 60}}
\put(475,740){\line( 0, 1){ 60}}
\put(470,795){\line( 1, 0){ 25}}
\put(510,750){\line( 1, 0){ 25}}
\put(600,795){\line( 1, 0){ 15}}
\put(640,750){\line( 1, 0){ 20}}
\put(166,709){
}
\put( 54,698){
}
\put(425,709){
}
\put(384,727){
}
\put( 85,800){\makebox(0,0)[lb]{\scriptsize $\mathrm{I}_m$}}
\put(290,695){\makebox(0,0)[lb]{$C_0$}}
\put(255,705){\makebox(0,0)[lb]{\scriptsize $\mathrm{II}^*$}}
\put(400,795){\makebox(0,0)[lb]{\scriptsize $\mathrm{I}_1$}}
\put(435,760){\makebox(0,0)[lb]{\scriptsize $\mathrm{I}_2$}}
\put(445,735){\makebox(0,0)[lb]{\scriptsize $\mathrm{I}_3$}}
\put(480,765){\makebox(0,0)[lb]{\scriptsize $\mathrm{I}_4$}}
\end{picture}
\end{center}
  \caption{$E_8$ instantons on a cyclic quotient.}
  \label{fig:E8Am}
\end{figure}
\fi

Taking into account the other $E_8$ from the line $C_\infty$ we obtain
the following
\begin{result}
  A collection of $k$ point-like $E_8$ instantons
on a $\C^2/\Z_m$ (that is, type $\mathsf{A}_{m-1}$) quotient singularity,
where $k\geq2m$,
yields $n_T'=k$ and local contribution to the gauge algebra
\begin{equation}
  \Gloc\cong\su(2)\oplus\su(3)\oplus\ldots\oplus\su(m-1)\oplus
     \su(m)^{\oplus(k-2m+1)}\oplus\su(m-1)\oplus\ldots\oplus\su(2).
\end{equation}
\end{result}
One may show that the case $k<2m$ reduces to the case
obtained by replacing $m$ with the integer part of $k/2$.

It is amusing to observe that this gauge algebra corresponds to the
semisimple part of the algebra found by Hanany and Witten \cite{HW:3d} in
three dimensions as the mirror of $U(m)$ gauge theory with $k$ flavors.
(The Hanany--Witten algebra is not semisimple, and contains $\gu(j)$'s in
place of our $\su(j)$'s.)  It would be interesting to find an explanation
of this fact, perhaps by compactifying our models to three dimensions.

It is satisfying to note that the quotient singularity $\C^2/\Z_3$ in
$S_H$ yields the same physics whether it is produced by a fibre of type
$\mathrm{I}_3$ or of type $\mathrm{IV}$, as discussed earlier.

If we slightly modify (\ref{eq:E8Am}) to
\begin{equation}
\begin{split}
a &= 3s^4t^2(-1+t^m)\\
b &= 2s^5t^3(s+t^\ell)\\
\delta &= 108s^{10}t^{6+m}(3s^2 - 3s^2t^m + s^2t^{2m} + 2st^{\ell-m} +
t^{2\ell-m}),
\end{split}  \label{eq:E8Dm}
\end{equation}
we produce a vertical line of $\Ist m$ fibres and hence a $\mathsf{D}_{m+4}$
quotient singularity in $S_H$. This collision contains $\ell+3$ instantons.
The resolution in this case starts exactly
as in figure \ref{fig:E8Am} except that every  $\mathrm{I}_n$ fibre
in the chain is replaced by an $\Ist n$ fibre. To complete the
resolution, every intersection of a $\Ist{n_1}$ fibre with a
$\Ist{n_2}$ fibre must be blown up to produce a curve of
$\mathrm{I}_{n_1+n_2}$ fibres. The remaining $\Ist0-\mathrm{II}^*$
intersection also requires a few blowups as we have already discussed
earlier. Taking into account the various
monodromies acting we obtain the following
\begin{result}
  A collection of $k$ point-like $E_8$ instantons
on a $\C^2/\mathsf{D}_{m+4}$ quotient singularity,
where $k=2m+6$ and $m>0$, produces $n_T'=2k-6$ and

\begin{multline}
  \Gloc\cong\su(2)\oplus\mf{g}_2\oplus\so(9)\oplus\so(3)\oplus
  \so(11)\oplus\so(5)\oplus\ldots\oplus\so(2m+5)\oplus
  \so(2m-1)\\ \oplus\so(2m+7)\oplus\so(2m-1)\oplus\ldots
  \oplus\so(9)\oplus\mf{g}_2\oplus\su(2),
\end{multline}
or, if $k\geq 2m+7$ and $m>0$, we have $n_T'=2k-6$ and
\begin{multline}
  \Gloc\cong\su(2)\oplus\mf{g}_2\oplus\so(9)\oplus\so(3)\oplus
  \so(11)\oplus\so(5)\oplus\ldots\oplus\so(2m+5)\oplus
  \so(2m-1)\\ \oplus\so(2m+7)\oplus\sp(m)\oplus\bigl(\so(2m+8)\oplus
  \sp(m)\bigr)^{\oplus(k-2m-7)}\oplus\so(2m+7)\\ \oplus\so(2m-1)\oplus\ldots
  \oplus\so(9)\oplus\mf{g}_2\oplus\su(2).
\end{multline}
\end{result}

The $m=0$ case was covered in section \ref{ss:J=0}.
If $6\leq
k<2m+6$ then replace $m$ by the integer part of
$k/2-3$. The $k=4$ or 5 cases reduce to the $A_2$ case in table
$\ref{tab:E81}$ and, as always, $k<4$ is trivial.

Finally we need to deal with the $\mathsf{E}_7$ singularity. We may put
\begin{equation}
\begin{split}
a &= s^4t^3\\
b &= s^5t^5(s+t^\ell)\\
\delta &= s^{10}t^9(4s^2 + 27(s+t^\ell)^2t).
\end{split}  \label{eq:E8E7}
\end{equation}
which gives a vertical line of $\mathrm{III}^*$ fibres and uses
$5+\ell$ instantons.

After some work we obtain the following
\begin{result}
  A collection of $k$ point-like $E_8$ instantons
on a $\C^2/\mathsf{E}_7$ quotient singularity, where $k\geq10$,
produces $n_T'=6k-40$ and
\begin{multline}
  \Gloc\cong\su(2)\oplus\mf{g}_2\oplus\mf{f}_4\oplus\mf{g}_2
  \oplus\su(2)\oplus\mf{e}_7\oplus\bigl(\su(2)\oplus\so(7)\oplus\su(2)
  \oplus\mf{e}_7\bigr)^{\oplus(k-10)}\\
  \oplus\su(2)\oplus\mf{g}_2\oplus\mf{f}_4\oplus\mf{g}_2
  \oplus\su(2).
\end{multline}
\end{result}
The cases where $k<10$ coincide with those of the $\mathsf{E}_8$
quotient singularity given in table \ref{tab:E8low}.


\section{The $\spnh$ Heterotic String on a K3 Surface}   \label{s:D}

The point-like instantons of the $\spnh$ heterotic string are
completely different and, in many ways, a little easier to analyze
than the $E_8$ instantons of the previous section. There are two types
of point-like instanton in the $\spnh$ case which have trivial
holonomy. First there is the
``simple'' instanton of Witten \cite{W:small-i}. Secondly there is the
hidden obstructer \cite{me:sppt}. This latter object has instanton
number 4 and sits at a quotient singularity which is at least as bad as
$\C^2/\Z_2$. The hidden obstructer has a massless tensor which leads
to a Coulomb branch in the moduli space. If we follow this branch
in the moduli space we can turn a hidden obstructer into a collection
of four simple instantons sitting on the same quotient singularity. We
will therefore restrict our attention in this section to only
considering simple instantons. In every case that a massless tensor
appears, we may then take note that it is possible to replace four simple
instantons by a hidden obstructer if we wish to do so.

We want to consider $X$ in the Weierstrass form
\begin{equation}
  y^2 = x^3 + p(s,t)\, x^2 + \varepsilon(t)\,x,
\end{equation}
where $p(s,t)$ is a cubic equation in $s$. This will put a line of
$\Ist{12}$ fibres along $s=\infty$ and give us at least two sections
which will lead to an unbroken perturbative gauge group of $\spnh$. We
will denote the line $s=\infty$ by $C_\infty$. We may relate this form
to the usual Weierstrass form by
\begin{equation}
\begin{split}
a &= \varepsilon - \ff13p^2\\
b &= \ff13p(\ff29p^2-\varepsilon)\\
\delta &= \varepsilon^2(4\varepsilon-p^2).
\end{split}   \label{eq:Dsim}
\end{equation}

As explained in \cite{AG:sp32}, a simple instanton is then associated
to a root of
$\varepsilon(t)$. If everything else is generic, this produces a
``vertical'' (i.e., $t$ is constant)
line of $\mathrm{I}_2$ fibres which gives an $\sp(1)$ nonperturbative
gauge symmetry. A zero of order $k$ in $\varepsilon(t)$ produces a
vertical line of $\mathrm{I}_{2k}$ fibres which, thanks to monodromy,
produces a gauge symmetry of $\sp(k)$.

If $\varepsilon(t)$ controls the location of the simple instantons, it
is clear that $p(s,t)$ must control the heterotic K3 moduli. In section
\ref{s:stab} we saw that the heterotic elliptic curve is given by a
double cover of $\P^1$ branched at four points. These four points are
located at $s=\infty$ and the three roots of $p(s)=0$. Now when we let this
elliptic curve vary in a family by varying $t$, we build an elliptic
surface. We can naturally describe the surface, $S_H$, in Weierstrass
form
\begin{equation}
  u^2 = p(s,t),   \label{eq:SHe}
\end{equation}
since $p$ is cubic in $s$.

To avoid hidden obstructers we choose $\Thet=\HS 4$ and put
$C_\infty$ along the isolated section \cite{me:sppt}. One can then
show that the
discriminant of $p(s,t)$ with respect to $s$ is a degree 24 polynomial
in $t$. This shows that (\ref{eq:SHe}) does indeed describe a K3 surface.

We now know exactly how to put singularities into $S_H$. For example,
we may make a $\C^2/\Z_n$ singularity by locally defining
\begin{equation}
  p(s,t) = s^3 -3s + (2+t^n),
\end{equation}
as this puts an $\mathrm{I}_{n}$ fibre into $S_H$ at $t=0$. If we substitute
this into (\ref{eq:Dsim}) then we generically do not worsen the
discriminant locus of the elliptic fibration on $X$. Thus we recover
the result that a $\C^2/\Z_n$ singularity in $S_H$ will not produce
any interesting nonperturbative physics. The same is true for any
other quotient singularity.

Interesting things do happen however when we let $t=0$ coincide with a
zero of $\varepsilon(t)=0$. Clearly this corresponds to a collision of
simple instantons with a quotient singularity.

Any time we have a singularity in $S_H$, $p(s,t)$ will have a zero
of total degree $\geq2$ somewhere (e.g., at $(s,t)=(1,0)$ in the above
$\C^2/\Z_n$ example). Suppose $\varepsilon$ is of degree $\geq4$ at
the same value of $t$. Now we see that the total degrees of
$(a,b,\delta)$ in (\ref{eq:Dsim}) are at least $(4,6,12)$. This means that
we must blow up the base (and correct for the non-minimality of the
Weierstrass model) as we have discussed above. Many blowups may be
required before $X$ becomes smooth, depending on how many instantons we have
and what quotient singularity appears in $S_H$.

Having written $S_H$ in Weierstrass form there is actually a rather
pretty correspondence between the way $X$ is blown up and the way that
the quotient singularity in $S_H$ is blown up which we will now
derive. This makes calculating the nonperturbative physics of simple
instantons on quotient singularities only a little harder than knowing
how to blow up the quotient singularities themselves.

Let us recall how to blow up $\mathsf{ADE}$ singularities in an elliptic
surface, $S_H$. The Weierstrass form (\ref{eq:SHe}) depicts the elliptic
fibre as a double cover of the $s$-line branched over the four points
given by $s=\infty$ and the 3 roots of $p(s,t)=0$ for a given $t$ in
the base. $S_H$ is then singular if and only if the branch locus
$p(s,t)=0$ is singular. The blowup of $S_H$ is therefore achieved by
blowing up the branch locus $p(s,t)=0$ in the $(s,t)$-plane until it
is smooth. For an example see the appendix of \cite{me:sppt}, and see
\cite{Mir:fibr} for a table of the resulting branch locus in every case.

So long as there are enough instantons on the singularity, each such
blowup in the $(s,t)$-plane maps directly to a blowup in the base of
the elliptic fibration on $X$. If there are too few instantons at any given
stage, $X$ may be
smooth before $S_H$ is. The further blowups for $S_H$ then produce no
new nonperturbative physics. The extreme case is when we have no
instantons on the quotient singularity in which case $X$ requires no
blowups in the base at all.

As the branch locus of the elliptic surface, $S_H$, is blown up, at
each stage the exceptional divisor may, or may not, be in the new
branch locus. The rule is clear: if the total degree of the branch
locus was even at the point which was blown up then the exceptional
divisor {\em will not\/} be in the branch locus and if the total degree
of the branch
locus was odd at the point which was blown up then the exceptional
divisor {\em will\/} be in the branch locus. This copies over into a
simple rule for how the blowup proceeds in $X$. The only cases we
actually need are ones in which $p(s,t)$ is of total degree 2 or 3 at
any stage. If $p(s,t)$ is of degree 2 then $a$ and $b$ will be of
degree 4 and 6 respectively (assuming we have enough
instantons). Doing the blowup and adjusting $L$ to preserve $K_X=0$,
we see that $a$ and $b$ will both be of degree 0 along the exceptional
divisor. Thus we introduce a line of $\mathrm{I}_n$ fibres for some
$n$. If, on the other hand, $p(s,t)$ is of degree 3, it is clear that
we introduce a line of $\Ist n$ fibres.

\iffigs
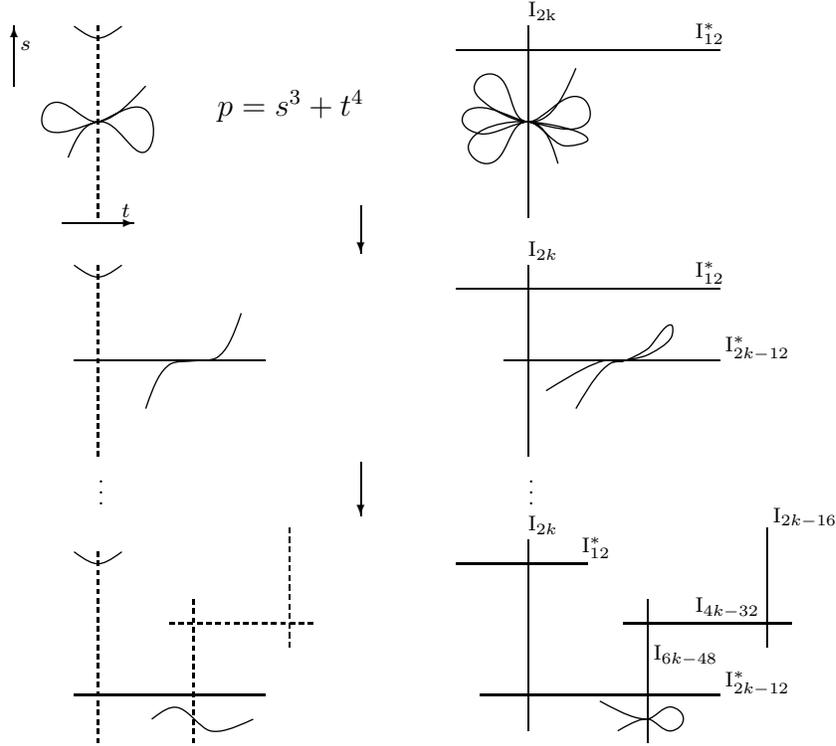
\begin{figure}[tbh]
\begin{center}
\setlength{\unitlength}{0.006250in}%
\begin{picture}(650,619)(10,200)
\thinlines
\multiput( 80,800)(0.00000,-7.80488){21}{\line( 0,-1){  3.902}}
\put( 60,800){
}
\put(120,750){
}
\put(440,800){\line( 0,-1){160}}
\put(380,780){\line( 1, 0){220}}
\put(480,765){
}
\put(580,785){\makebox(0,0)[lb]{\scriptsize$\Ist{12}$}}
\put(440,805){\makebox(0,0)[lb]{\scriptsize$\mathrm{I_{2k}}$}}
\multiput( 80,600)(0.00000,-7.80488){21}{\line( 0,-1){  3.902}}
\put( 60,520){\line( 1, 0){160}}
\put(440,600){\line( 0,-1){160}}
\put(380,580){\line( 1, 0){220}}
\put(420,520){\line( 1, 0){180}}
\put(300,650){\vector( 0,-1){ 40}}
\put(300,435){\vector( 0,-1){ 45}}
\multiput( 80,360)(0.00000,-7.80488){21}{\line( 0,-1){  3.902}}
\put( 60,240){\line( 1, 0){160}}
\multiput(160,200)(0.00000,7.74194){16}{\line( 0, 1){  3.871}}
\multiput(140,300)(7.74194,0.00000){16}{\line( 1, 0){  3.871}}
\multiput(240,280)(0.00000,8.00000){13}{\line( 0, 1){  4.000}}
\put(440,370){\line( 0,-1){160}}
\put(380,350){\line( 1, 0){110}}
\put(400,240){\line( 1, 0){200}}
\put(540,200){\line( 0, 1){120}}
\put(520,300){\line( 1, 0){140}}
\put(640,280){\line( 0, 1){100}}
\put( 10,750){\vector( 0, 1){ 50}}
\put( 50,635){\vector( 1, 0){ 60}}
\put( 60,600){
}
\put(120,480){
}
\put( 60,360){
}
\put(125,220){
}
\put(500,235){
}
\put(480,480){
}
\put(180,720){\makebox(0,0)[lb]{$p=s^3+t^4$}}
\put(580,585){\makebox(0,0)[lb]{\scriptsize$\Ist{12}$}}
\put(440,605){\makebox(0,0)[lb]{\scriptsize$\mathrm{I}_{2k}$}}
\put( 80,400){\makebox(0,0)[lb]{\scriptsize$\vdots$}}
\put(440,400){\makebox(0,0)[lb]{\scriptsize$\vdots$}}
\put(605,520){\makebox(0,0)[lb]{\scriptsize$\Ist{2k-12}$}}
\put(440,375){\makebox(0,0)[lb]{\scriptsize$\mathrm{I}_{2k}$}}
\put(485,355){\makebox(0,0)[lb]{\scriptsize$\Ist{12}$}}
\put(605,240){\makebox(0,0)[lb]{\scriptsize$\Ist{2k-12}$}}
\put(580,305){\makebox(0,0)[lb]{\scriptsize$\mathrm{I}_{4k-32}$}}
\put(645,380){\makebox(0,0)[lb]{\scriptsize$\mathrm{I}_{2k-16}$}}
\put(545,265){\makebox(0,0)[lb]{\scriptsize$\mathrm{I}_{6k-48}$}}
\put( 15,780){\makebox(0,0)[lb]{\scriptsize$s$}}
\put(100,640){\makebox(0,0)[lb]{\scriptsize$t$}}
\end{picture}
\end{center}
  \caption{$k$ simple instantons on an $\mathsf{E}_6$ singularity.}
  \label{fig:D8E6}
\end{figure}
\fi

Having blown up the branch locus for $S_H$, the exceptional divisor
within $S_H$ may be read off as the corresponding double cover of the
blowup of the branch locus. Any exceptional curve which was found to
be in the branch locus will of course end up as a rational curve in
$S_H$. Any exceptional curve which was not left in the branch locus
will be covered twice to appear in $S_H$. There are two
possibilities. If this curve was completely disjoint from the branch
locus then it will appear twice in $S_H$. If it intersected the branch
locus twice, it will appear as a single curve in $S_H$.

These rules map into the blowup of $X$ as follows. The exceptional
curves in the branch locus map to $\Ist n$ fibres without monodromy and
thus give $\so(2d)$ gauge algebras.
The exceptional curves not in the branch
locus will give $\mathrm{I}_n$ fibres and the fact that the do not hit
the branch locus means they will not intersect curves of $\Ist n$
fibres. This can be shown to imply that they have no monodromy and
thus lead to $\su(n)$ gauge algebras. The other exceptional curves not
contained in the branch locus but intersecting it lead to $\sp(n)$
gauge algebras.

This gives all the rules required to determine our problem. In figure
\ref{fig:D8E6} we give an example of the picture of $k$ simple instantons on an
$\mathsf{E}_6$ singularity. On the left we show the way that the branch locus
is resolved. The dotted lines are curves not in the branch locus and
solid lines are curves which are in the branch locus. These are the
same sorts of pictures as appear in \cite{Mir:fibr} and the appendix of
\cite{me:sppt}. On the right we show the corresponding discriminant
for $X$. As always, the curly lines are curves of $\mathrm{I}_1$ fibres.

\begin{table}[t]
\renewcommand{\arraystretch}{1.5}
$$
\begin{array}{|c|c|c|l|}
\hline
&k_{\mathrm{min}}&n_T&\multicolumn{1}{c|}{\Gloc}\\
\hline
\mathsf{A}_{m-1},\;m\text{ even}&2m&\frac{m}2&
{\renewcommand{\arraystretch}{1.0}\arraycolsep=0pt
\begin{array}[t]{l}
\sp(k)\oplus\su(2k-8)\oplus\su(2k-16)\oplus\ldots\\
\quad\ldots\oplus\su(2k-4m+8)\oplus\sp(k-2m)\end{array}}\\
\mathsf{A}_{m-1},\;m\text{ odd}&2m-2&\frac{m-1}2&
{\renewcommand{\arraystretch}{1.0}\arraycolsep=0pt
\begin{array}[t]{l}
\sp(k)\oplus\su(2k-8)\oplus\su(2k-16)\oplus\ldots\\
\quad\ldots\oplus\su(2k-4m+4)\end{array}}\\
\mathsf{D}_{m+4},\;m\text{ even}&2m+8&m+4&
{\renewcommand{\arraystretch}{1.0}\arraycolsep=0pt
\begin{array}[t]{l}
\sp(k)\oplus\sp(k-8)\oplus\so(4k-16)\oplus\sp(2k-16)\\ \quad
\oplus\so(4k-32)\oplus\sp(2k-24)\oplus\so(4k-48)\oplus\ldots\\
\qquad\ldots\oplus\sp(2k-4m-8)\oplus\so(4k-8m-16)\\ \quad\qquad\oplus
\sp(k-2m-8)^{\oplus2}\end{array}}\\
\mathsf{D}_{m+4},\;m\text{ odd}&2m+6&m+3&
{\renewcommand{\arraystretch}{1.0}\arraycolsep=0pt
\begin{array}[t]{l}
\sp(k)\oplus\sp(k-8)\oplus\so(4k-16)\oplus\sp(2k-16)\\ \quad
\oplus\so(4k-32)\oplus\sp(2k-24)\oplus\so(4k-48)\oplus\ldots\\
\qquad\ldots\oplus\sp(2k-4m-4)\oplus\so(4k-8m-8)\\
\qquad\oplus\sp(2k-4m-12)\oplus\su(2k-4m-12)\end{array}}\\
\mathsf{E}_6&8&4&
{\renewcommand{\arraystretch}{1.0}\arraycolsep=0pt
\begin{array}[t]{l}
\sp(k)\oplus\so(4k-16)\oplus\sp(3k-24)\oplus\su(4k-32)\\
\quad\oplus\su(2k-16)\end{array}}\\
\mathsf{E}_7&12&7&
{\renewcommand{\arraystretch}{1.0}\arraycolsep=0pt
\begin{array}[t]{l}
\sp(k)\oplus\so(4k-16)\oplus\sp(3k-24)\oplus\so(8k-64)\\
\quad\oplus\sp(2k-20)\oplus\sp(3k-28)\oplus\so(4k-32)\\ \qquad
\oplus\sp(k-12)\end{array}}\\
\mathsf{E}_8&11&8&
{\renewcommand{\arraystretch}{1.0}\arraycolsep=0pt
\begin{array}[t]{l}
\sp(k)\oplus\so(4k-16)\oplus\sp(3k-24)\oplus\so(8k-64)\\
\quad\oplus\sp(5k-48)\oplus\so(12k-112)\oplus\sp(3k-32)\\
\qquad\oplus\sp(4k-40)\oplus\so(4k-32)\end{array}}\\
\hline
\end{array}
$$
\caption{$k$ simple instantons on an $\mathsf{ADE}$ singularity.}
\label{tab:gensim}
\end{table}

In table \ref{tab:gensim} we show the complete results for all of the
quotient singularities. This time the $24-k$ instantons away from the
singularity will produce their own nonperturbative gauge algebra. If
we assume all are disjoint from each other the total gauge algebra is
given by
\begin{equation}
  \cG \cong \so(32)\oplus\sp(1)^{\oplus(24-k)}\oplus\Gloc.
\end{equation}
In each case we have imposed a minimum number, $k_{\mathrm{min}}$ on
the number of instantons.
The rules for a
smaller number of instantons can be fairly involved and are best
treated case by case. In this event the number of massless tensors can be
less than that of table \ref{tab:gensim} and some of the factors in
the gauge symmetry can be missing or, sometimes, different.

The case of $\C^2/\Z_2$
(i.e., $A_1$) corresponds to $\Gloc=\sp(k)\oplus\sp(k-4)$
and had already been determined in \cite{me:sppt}. The other cyclic
groups had previously been conjectured in \cite{In:RG6} and our
results agree up to abelian groups, which we are ignoring. The case
of $D_4$ can be seen to correspond to the case of two coalescing
hidden obstructers discussed in \cite{me:sppt}.


\section{Equivalences}   \label{s:equiv}

At first sight our results for the $E_8$ point-like instantons and the
simple $\spnh$ instantons appear to be quite different. It is
well-known however that the two heterotic strings are mapped to each
other fairly easily if we compactify on a circle (or a torus). We may
use this fact to find a connection between the two sets of results.

Suppose we specify the data of what kind of singularities the K3 surface $S_H$
has and how the point-like instantons are positioned with respect to the
singularities, and work in the Coulomb branch where nonzero expectation
values have been given to massless tensors.  If the point-like instantons
are $E_8$ instantons, let
the corresponding F-theory be compactified on the \CY\ threefold
$X_1$. If the point-like instantons are simple $\spnh$ instantons, let
the corresponding F-theory be compactified on the \CY\ threefold
$X_2$. We will now show that $X_1$ is birationally equivalent to
$X_2$.

Let $X$ refer to either $X_1$ or $X_2$. Let us recall the steps in
showing that $X$ is an elliptic
fibration with a section. We refer to
\cite{AL:ubiq,MV:F,AG:sp32,me:lK3} for details. Let a
type IIA string on $X$ be dual to a heterotic string on $S_H\times
T^2$. With a few caveats this implies that $X$ is a K3-fibration. The
moduli space of $T^2$ together with its vector bundle (i.e., Wilson
lines) is part of the moduli space of complexified K\"ahler forms on
$X$. The size of the $T^2$ itself may be mapped to part of the
K\"ahler form data given by the generic K3 fibre of $X$.

One can argue that this correspondence shows that the Picard lattice of
the generic K3 fibre contains the unimodular lattice $\Gamma_{1,1}$
and that this lattice is monodromy invariant in the fibration. This
may be used to deduce the fact that this generic K3 fibre is elliptic
with a section and that the elliptic fibre and the section are both
monodromy-invariant as homology classes. This then shows that $X$ is
an elliptic fibration with a (birational) section.

Let us consider the $E_8\times E_8$ heterotic string compactified such
that all the instantons are point-like. In this case we know that the
perturbative part of the gauge group contains
$E_8\times E_8$. The Cartan lattice of this group must be part of the Picard
lattice of
the generic K3 fibre. As such, one may now show that this Picard
lattice contains the unimodular lattice $\Gamma_{1,17}$. Now
$\Gamma_{1,17}$ may be decomposed as $\Gamma_{1,1}\oplus\Lambda$, where
$\Lambda$ is a unimodular definite rank 16 lattice, in two different
ways. $\Lambda$ is the lattice associated with $E_8\times E_8$ or
$\spnh$. This leads to two different elliptic fibrations (with
section) of this generic K3 fibre, one of which will produce the
$E_8\times E_8$ perturbative gauge symmetry and the other of which
will produce an $\spnh$ perturbative gauge symmetry when we go to the
F-theory limit by decompactifying the $T^2$.

In other words we may start with one heterotic string on a K3 surface,
compactify further on a 2-torus and then decompactify along a
2-torus in a different way to obtain the other heterotic string. This process
involves manipulating the K\"ahler form on $X$ and we may very well
produce flops in $X$ in the process. Note however that we have not
touched the complex structure of $X$. The complex structure data of
$X$ controls the hypermultiplets which specify the moduli of $S_H$ and
the instanton locations. We have therefore shown that $X_1$ is
birationally equivalent to $X_2$.

It is worth emphasizing that the point-like instantons played an
important r\^ole here. In a more generic case where part of the
primordial gauge group is broken, we may not see $\Gamma_{1,17}$ as
part of the Picard lattice of the generic K3 fibre and the above
argument need not hold.

The fact that $X_1$ and $X_2$ are birationally equivalent implies that
their Hodge numbers are the same. We know that
\begin{equation}
  h^{1,1}(X) = \rank(\cG)+n_T+3.   \label{eq:h11}
\end{equation}
This leads to relations between the $E_8\times E_8$ and $\spnh$ cases
that we may check. For example, let us take $k$ instantons on a
$\C^2/\Z_m$ quotient singularity where $m$ is even and $k\geq2m$. In
both cases---either $E_8$ instantons or simple $\spnh$ instantons---we
find that
\begin{equation}
  h^{1,1}(X) = 44+km-m^2+k,\label{eq:h11Am}
\end{equation}
in agreement with our assertion. The rank of the gauge groups and the
number of massless tensors is different in each case however---it is
only their sum which is the same. The two elliptic fibrations on $X$
are very different.

At this point we should mention the r\^ole of the \MW\
group. If this group has nonzero rank then we expect possible $U(1)$
factors in the gauge group which we have ignored up to now. Such
effects would contribute to (\ref{eq:h11}) and would have invalidated
(\ref{eq:h11Am}). When one analyzes the $\spnh$ instantons in terms of
open strings one often finds nonperturbative gauge groups of the form
$U(n)$. In many cases the $U(1)$ symmetries are broken due to
consideration of anomalies
\cite{BLPSSW:so32}. In \cite{In:RG6} Intriligator found abelian groups
when he analyzed the case of instantons on a $\C^2/\Z_m$ quotient
singularity.
To find the agreement above we assumed that {\em all\/} of these
$U(1)$'s are broken. Of course, there may be conspiracy which gives an
equal nonzero rank to the \MW\ group for both $X_1$ and $X_2$ but this
seems rather unlikely in general. This point should be investigated further.


\section{Discussion}   \label{s:conc}

We have considered the heterotic string on a K3 surface, $S_H$, in terms of
F-theory on a \CY\ threefold $X$. By taking the large volume limit of
the K3 surface we have produced a stable degeneration of $X$. In this
stable degeneration the moduli of $X$ which control the moduli of
$S_H$ are nicely separated from the moduli which control the vector
bundle structure on $S_H$. This allows the F-theory moduli space to be
mapped explicitly to the heterotic moduli space.

We then analyzed point-like instantons in the heterotic string. The
reader may have noticed that the stable degenerations in section
\ref{s:stab} were analyzed in terms of a family which was precisely
the local description of a point-like instanton in both the $E_8\times
E_8$ and the $\spnh$ case. That is, these stable degenerations must be
intimately linked to the notion of a point-like instanton. Actually
it can be seen why this is so. The point-like instantons which we analyzed
are the only objects which correctly separate the moduli of
$S_H$ from its vector bundle.

Consider, for example, a point-like
instanton which does {\em not\/} have trivial holonomy. Such an example, with
local holonomy $\Z_2$, was discussed in \cite{BLPSSW:so32}. Topology
forces such an object to sit at a quotient singularity of $S_H$ as
only then is it surrounded by a lens space which allows its holonomy
to be exhibited. Therefore the bundle moduli (i.e., its location) are
tied to the moduli of $S_H$. Similarly the hidden obstructer of
\cite{me:sppt} is forced to be at a quotient singularity. Giving an
instanton nonzero
 size only makes the mixing of moduli worse. Therefore
we should not be surprised that the stable degenerations considered
are so closely tied to the point-like instantons we analyzed.

Let us now summarize a few features of the rules for the coalesced
instantons on the quotient singularities. We call the simple $\spnh$
instanton a simple instanton for brevity.
\begin{itemize}
\item[1.] A quotient singularity without point-like instantons produces no
interesting nonperturbative physics.
\item[2.] Fewer than four instantons on a quotient singularity produce no
nonperturbative physics (beyond that
of a smooth point) for both the simple and $E_8$ cases.
\item[3.]
In the case of $k$ $E_8$ instantons on any smooth or singular point we
always have $n_T'\geq k$.
\item[]
In the case of $k$ simple instantons on any smooth or singular point we
always have $\sp(k)$ as a factor of $\Gloc$.
\item[4.]
Four $E_8$ instantons on any quotient singularity gives
$\Gloc\cong\su(2)$ and $n_T'=4$
\item[]
Four simple instantons on any quotient singularity gives
$\Gloc\cong\sp(4)$ and $n_T=1$.
\item[5.]
Increasing the number of $E_8$ instantons on a singularity
beyond a certain minimum number increases the number of terms in
$\Gloc$ but does not raise the rank of each term.
\item[]
Increasing the number of simple instantons on a singularity
beyond a certain minimum number does not increase the number of terms in
$\Gloc$ but does raise the rank of each term.
\end{itemize}

Consider the ``record'' gauge algebras we might make by
coalescing all 24 instantons
on an $\mathsf{E}_8$ singularity. In the $E_8$ case we find the gauge algebra
given by
\begin{equation}
  \cG \cong \mf{e}_8^{\oplus17}\oplus
\mf{f}_4^{\oplus16}\oplus
\mf{g}_2^{\oplus32}\oplus\su(2)^{\oplus32},  \label{eq:E8big}
\end{equation}
which is rank 296 and $n_T=192$. The F-theory \CY\ threefold
$X_{\mathrm{big}}$,
corresponding to this has $h^{1,1}=\rank(\cG)+n_T+3=491$ and $h^{2,1}$
given by the number of moduli minus 1. As we have only the 12
remaining deformations of the K3 after fixing the $\mathsf{E}_8$ singularity,
we have $h^{2,1}=11$. These are the Hodge numbers of the well-known
\CY\ threefold which holds the current record for largest Euler
characteristic. It is mirror to the \CY\ threefold {\em known\/} to have the
most negative Euler characteristic \cite{Gross:ft} for an elliptic
\CY\ threefold. Toric methods can be used to construct $X_{\mathrm{big}}$
and indeed the gauge group (\ref{eq:E8big}) results.\footnote{P.S.A.\
would like to thank M.~Gross for conversations on this point.}
This has also been explained in \cite{CPR:E8-17}. Playing the same
game for the simple $\spnh$ instantons we get a gauge group
\begin{multline}
\cG\cong\so(32)\oplus\sp(24)\oplus\so(80)\oplus\sp(48)\oplus\so(128)
\oplus\sp(72)\\ \oplus\so(176)\oplus\sp(40)\oplus\sp(56)\oplus\so(64),
\end{multline}
which is rank 480 and $n_T=8$. Again the Hodge numbers of the F-theory
\CY\ threefold are $h^{1,1}=491$ and $h^{2,1}=11$ in agreement with
our assertions in section \ref{s:equiv}.

One might speculate that this latter gauge symmetry is the largest one
may acquire in an $N=1$ compactified string theory in 6 dimensions. It
might be conceivable that one may push further by making $S_H$
Planck-sized. This would take it out of the realm of the F-theory
analysis we have done. However, one always seems to need instantons to make
nonperturbative gauge symmetries and we have used up all 24 in our
quotient singularity here. Thus the speculation may be correct.

It is worth noting that we can produce any simple gauge algebra
(below a certain rank) within $\Gloc$ by coalescing certain $E_8$
instantons at a singularity in $S_H$ whereas we cannot produce the
exceptional algebras from the simple $\spnh$ instantons. Since the
$E_8\times E_8$ heterotic string and the $\spnh$ heterotic string on a
K3 surface are actually expected to be the same thing, once the moduli
have been suitably reinterpreted, this must just be because we have not
probed the directions in moduli space given by the R-R fields.

In this paper we have largely ignored the moduli of the vector bundle in the
heterotic string by fixing all instantons to be point-like. One of the
virtues of the stable degeneration method we have used here is that it
gives a natural description of the moduli space of the vector bundle
as well as the underlying base space. This was described in
\cite{FMW:F} for the $E_8\times E_8$ string in terms of del Pezzo
surfaces. It would be interesting to extend this to the $\spnh$ case,
which seems clearly possible from our analysis here. One could then
explore more fully the moduli space of heterotic string on K3 surfaces.


\section*{Acknowledgements}

It is a pleasure to thank R.~Donagi, R.~Friedman, M.~Gross,
K.~Intriligator, S.~Kachru, E.~Silverstein, and E.~Witten for useful
conversations.  The work of P.S.A.\ is supported by DOE grant
DE-FG02-96ER40959.  The work of D.R.M.\ is supported in part by the
Harmon Duncombe Foundation and by NSF grants DMS-9401447 and
DMS-9627351.

\section*{Added in proof:}

The results in this paper are dependent upon some assumptions about the
Ramond-Ramond moduli having been set to zero. This may raise some rather
subtle issues when computing the gauge algebra which results from
monodromy. In particular, although the geometry which leads to result 4 is
certainly correct, its interpretation in terms of gauge algebras may
require some modification. We thank K.~Intriligator for pointing out to us
that result 4 was potentially incorrect.


\end{document}